\documentclass[twocolumn]{aastex631}
\usepackage{amsmath,verbatim,rotating}
\usepackage{CJKutf8}
\newcommand{\cntext}[1]{\begin{CJK*}{UTF8}{gbsn}#1\end{CJK*}}
\allowdisplaybreaks

\usepackage{xcolor}

\newcommand{\tti}{t_{\rm TI}}
\newcommand{\tff}{t_{\rm ff}}
\newcommand{\rcs}{R_{\rm cs}}

\newcommand{\rjet}{r_{\rm jet}}
\newcommand{\hjet}{h_{\rm jet}}

\newcommand{\rb}{r_{\rm B}}
\newcommand{\rhl}{r_{\rm HL}}
\newcommand{\rdep}{R_{\rm reset}}
\newcommand{\rbhl}{r_{\rm BHL}}
\newcommand{\newtext}[1]{#1}

\shorttitle{A New Framework for Jet Feedback}
\shortauthors{L\"u \& Ricker}

\begin{document}

\title{A New Framework for AGN Accretion and Jet Feedback in Numerical Simulations}

\correspondingauthor{Ying-He L\"u}
\email{yl790@cam.ac.uk}

\author{Ying-He Celeste L\"{u} (\cntext{吕映荷})}
\affil{Cavendish Radio Astronomy Group, University of Cambridge, Cambridge CB3 0HE, UK}
\author{Paul M.~Ricker}
\affil{Department of Astronomy, National Center for Supercomputing Applications, and Illinois Center for Advanced Studies of the Universe, University of Illinois at Urbana-Champaign, Urbana, IL, 61801, USA}

\begin{abstract}

Accurate modeling of active galactic nucleus (AGN) feedback, especially due to relativistic jets, is crucial for understanding the cool-core problem in galaxy clusters. We present a new subgrid method to model accretion onto and feedback from AGN in hydrodynamical simulations of galaxy clusters. Instead of applying the traditional Bondi formalism, we use a sink particle algorithm in which the accretion flux is measured directly through a control surface. A weighting kernel is used to reset the gas properties within the accretion radius at the end of each timestep. We implement feedback in the form of bipolar jets whose properties are tied to the accretion rate. The method is tested with a spherically symmetric Bondi gas flow problem and a Bondi-Hoyle-Lyttleton wind problem, with and without jet feedback. We discuss the reliability of this model by comparing our jet simulations with those in the literature, and we examine the dependence of test results on parameters such as the resolution and size of the jet injection region. We find that the sink particle model can account for the $\alpha$ factor in accretion measurement, and the accretion radius must be resolved with at least two zones to produce realistic black hole accretion. We also show how under-resolving the AGN feedback region in simulations can impact the feedback energy deposited and the jet dynamics. The code described here is the framework for a feedback model, described in a companion paper, that will use accretion disk modeling to more self-consistently determine the feedback efficiency. 
\end{abstract}

%===============================================================================

%===============================================================================

\section{Introduction}

A number of physical processes shape the observational properties of the hot, X-ray emitting intracluster medium (ICM) in galaxy clusters. These include radiative cooling, cosmic rays, thermal conduction, viscosity, turbulence, and black hole jet heating. Observations have shown that the cold gas formation rate at the centers of some galaxy clusters is much lower than expected based on X-ray luminosity, a discrepancy called the ``cool-core problem'' \citep{Peterson03}. Cool-core clusters have high central densities and low central temperatures, corresponding to short cooling times, but generally exhibit $\sim 1\%$ of the star formation rate expected under these conditions \citep{McDonald2018}. Active galactic nuclei (AGN) are thought to be the major feedback source that prevents hot gas from cooling and reduces star formation \citep{McNamara, Harrison2024}. 

AGNs are among the most energetic objects in the Universe, characterized by extremely luminous cores powered by the infall of gas from a relativistic accretion disk onto a central supermassive black hole (SMBH). Two major feedback modes have been identified for AGN from observations \citep{Fabian}. The radiative (or quasar/wind) mode operates when the SMBH accretes at a rate close to the Eddington limit; the mechanical (or radio) mode, characterized by radio-emitting jets, operates at lower accretion rates. Mechanical feedback is generally thought to heat the ICM more efficiently, and synchrotron emission from relativistic jets has been observed at radio frequencies for bright X-ray sources \citep{Kolokythas}. The estimated jet power of AGN-inflated cavities is correlated with the X-ray luminosity within the cluster cores \citep{birzan}. The X-ray cavities also show correlation with radio emission; an example is the NGC 1275 in the Perseus cluster. The Phoenix cluster is a rare contrasting example in which AGN feedback does not appear to fully offset gas cooling \citep{McDonald}.
 
In addition to X-ray cavities and radio bubbles, optical, infrared, and submillimeter observations have revealed molecular gas filaments in a number of cool-core clusters. In H$\alpha$ and sometimes CO, observers have discovered radially extended cold filaments around the X-ray cavities \citep{Conselice, Simionescu, Vantyghem, Olivares}. 
Optical spectroscopy has shown that these filaments are ionized by shocks \citep{Boselli}. 
ALMA observations show that the filaments are entrained by the X-ray-emitting hot ICM, flowing inward toward the AGN and then rising alongside AGN jets and behind the bubbles they inflate \citep{McNamara2016,Tremblay2018}. In addition, larger-scale structures can be present, such as merger shocks and cold fronts, turbulence, and ripples produced by sound waves \citep{Fabian2017}.

Multiphase hydrodynamic simulations of galaxy clusters can now produce realistic feedback loops and reproduce some of the structures observed in cool-core clusters \citep[e.g.,][]{gaspari2017cca,Li, Ehlert}, and there has been significant progress in understanding different heating mechanisms and how jet feedback is coupled with them \citep[e.g.,][]{Yang2, Martizzi, Bourne2023}. However, due to the large range of spatial and temporal scales involved in the problem, it is infeasible to resolve length scales down to the size of the black hole for a cluster-scale simulation. As a result, accretion and feedback are usually included through a phenomenological subgrid prescription. As an example, in adaptive mesh refinement (AMR) simulations, the accretion rate onto the black hole is usually determined by assuming that the subgrid flow is described by the spherically symmetric Bondi accretion model \citep{Bondi}, with an arbitrary boost factor applied to the rate computed from densities and sound speeds measured on resolved scales, which may be considerably larger than the accretion radius \citep{Booth}. The energy input rate of the feedback is taken to be a constant fraction of this accretion rate, and an accretion rate-dependent switch decides between quasar-mode and mechanical-mode feedback \citep[e.g.,][]{Richardson,Meece}. In cosmological simulations, such as \newtext{the EAGLE \citep{Schaller}, Horizon-AGN \citep{Dubois}, IllustrisTNG and TNG-Cluster \citep{Nelson2019, Nelson2024}, Simba \citep{dave_etal_2019}, and Obsidian \citep{Rennehan2024} simulations}, where resolution is on the scale of kiloparsecs, such approximations are frequently used. In some cases, to account for cold and hot gas, the boost factor can vary for gas of different temperatures \citep[e.g.,][]{Steinborn}. The feedback is mostly imposed through isotropic energy injection into the cells surrounding the SMBH.

In all these simulations, ad hoc parameters are used to follow a non-self-consistent black hole accretion history. The assumption of spherical accretion is unrealistic, and the necessary boost factor is resolution-dependent. For example, \cite{Negri} have found that the approximate Bondi formula can lead to both over- and underestimation of BH growth, depending on resolution and on how variables entering the accretion rate equation are calculated. However, for a problem involving accretion and feedback, often no single separation scale exists; as resolution improves, the necessary subgrid physics varies. To accommodate these facts, many alternative accretion models have been investigated. For example, some groups have used different inputs to a Bondi-like equation, such as the large-scale properties of host galaxies \citep{Angles}. Others take into account some disk structure or angular momentum effects, but without aiming to actually follow the accretion disks themselves \citep{Power}. Studies of convergence and validation for these models are still limited.

A major limitation of these ``hot Bondi''-like accretion models is that they make no provision for thermal instability in the gas on scales between the resolution length element and the size of the accretion disk. 
Despite the fact that heating dramatically suppresses star formation in cool-core clusters, the H$\alpha$ filaments strongly indicate the presence of cool gas. \cite{McCourt}, by studying an ideal gravitationally stratified plasma, found that the non-linear saturation of thermal instability is governed by the ratio of thermal instability timescale $\tti$ to the dynamical (``free-fall") timescale $\tff$. When $\tti/\tff \leq 1$, the cold gas condenses, and the plasma develops an extended multiphase structure. For spherical systems, this criterion becomes $\tti/\tff \leq 10$. Further studies have confirmed that, when the ratio falls below this value, gas condensation occurs, and within a cluster potential it also precipitates into the region closer to the SMBH and is accreted by it \citep{Tremblay2016,Voit, Nobels}. The condensed clouds then cool and form a filamentary nebula structure, which can explain the H$\alpha$ filaments. This scenario has come to be known as ``chaotic cold accretion'' \citep[CCA;][]{Gaspari2013}. This discovery has motivated a way to treat accretion as an inflow through a spherical region. For example, \cite{Gaspari2013} evacuated the gas within an accretion sphere in their simulations and used the mass of gas lost per timestep as the accretion rate for that step. Given that the flow becomes mostly supersonic at parsec scales, a complete evacuation is justified. However, given the turbulent and multiphase ICM structure involved, the exact kinematics of the sink-region gas is more complicated and involves both inflow and outflow. Methods that average density over a few large zones surrounding the SMBH do not consider these effects, and are thus inaccurate compared to a flux-based method that measures the actual mass inflow.

The notion of ``sink particles'' was first used in star formation studies. They were developed with the general aim of coping with regions of a flow that accrete infalling material but whose internal structure is unresolved. Sink particles were introduced by smoothed particle hydrodynamics (SPH) codes \citep{Bate} as absorbers that accrete other particles that approach within a certain distance and meet various other criteria. \cite{Krumholz} were the first to introduce sink particles in the Eulerian, grid-based code ORION, built upon the AMR technique. In addition to determining the accretion rate, the gas mass accreted in a timestep, $\dot{M} \Delta t$, is also removed from the same region along with the related momentum and total energy. Later \cite{Federrath} introduced a sink particle implementation into the FLASH code \citep{Flash,Flash3} with an approach that deviates considerably from the original \citep{Krumholz} method; they tested whether gas is gravitationally bound to the accretor before adding it to the accretor's mass. Since then a number of implementations have been developed in various other grid-based codes, such as Enzo \citep{Wang},  RAMSES \citep{Bleuler}, and Athena \citep{Gong}. Most of these works incorporate various methods to calculate the mass flux and a further correction of the accretion region to smooth out numerical effects. These sinks provide a realistic way to model accretion without damaging the rest of the calculation on the grid. 

In the absence of a detailed understanding of jet and wind launching from accretion disks, the methods for introducing feedback linked to accretion have typically also relied on adjustable parameters. The quasar-/radio-mode switch is one; others include the energetic feedback efficiency and the mass-loading factor. However, relativistic jet launching and propagation have been studied via (magneto)hydrodynamical simulation for decades \citep[see the review by][]{Marti}.
Most recently, general relativistic magnetohydrodynamics (GRMHD) simulations have been able to generate jets self-consistently from rotating magnetized accretion flows \citep{McKinney2009}. 
These successes enable us to consider unifying simulation results on accretion disk scales with larger scales associated with thermal instability in cluster cores \citep{Gaspari2017}.
Effects of heating and feedback from relativistic jets on kiloparsec scales have been studied by \cite{Mukherjee} and \cite{Perucho}, and updated physics with accretion disk structure and disk winds has been considered \citep[e.g.,][]{Yuan2018, Costa}. Recent efforts including \cite{Talbot2021}\newtext{, \cite{Husko2023},} and \cite{Rennehan2024}, incorporate varying accretion disk configurations into a unified model.

Several major questions concerning the details of AGN feedback and its interplay with the ICM have so far remained unanswered. In particular, how does the energy in AGN jets couple to the ICM? Why do some clusters have cool cores while others do not? What processes govern transitions between the two states? How do we explain objects such as the Phoenix cluster? A better understanding of these questions requires more knowledge of the plasma physics governing diffusive processes in the ICM \citep[e.g.,][]{Yang}. However, to address them, a more physically motivated subgrid model for AGN accretion and feedback is also required. 

Motivated by these unresolved questions and previous models, we have developed a new subgrid framework for accretion and jet feedback in hydrodynamic simulations of AGN feedback using FLASH. The accretion model resembles a sink particle algorithm, measuring accretion onto the SMBH via numerical fluxes across a spherical surface instead of relying on the Bondi solution. With this formulation we further investigate different ways to inject jets and characterize mechanical feedback, and we compare the jet dynamics and feedback properties with other smaller-scale simulations to test the convergence of our model. Upon this framework, we are also building a feedback model that incorporates the evolution of the AGN accretion disk, to be described in a companion paper (Khan et al.\ 2024, in preparation, hereafter Paper~II).

The paper is organized as follows. In \S~2 we introduce the details of our accretion method and discuss convergence tests in the absence of feedback. In \S~3 we describe the jet feedback model and examine the effects of different parameters on feedback. In \S~4 we briefly survey the literature on jets in numerical simulations at different scales and compare our test results with previous work. We offer conclusions in \S~5.
\\

\section{Accretion Method and Tests}

For all our simulations, we use the adaptive mesh refinement code FLASH 4 with its directionally unsplit hydrodynamic solver and a fixed gravitational potential \citep{Flash,Flash3}. AGN are treated as particles in our simulations. For the runs in this paper we place one AGN particle at the center of the grid at rest to approximate an isolated cluster environment, but all the accretion and feedback algorithms are connected to properties of the AGN particle in the particle's rest frame and thus can be used in cosmological simulations with multiple AGN particles. 

\subsection{Sink Particle Algorithm}

In most previous work involving sink particles, some form of reset of the gas properties within the sink particle radius is performed to conserve mass, energy and momentum. The reset usually involves an arbitrary kernel function and some correction parameters. Motivated by this work, we describe our model as follows.

\subsubsection{Measuring the Accretion Rate}

In our model, the accretion configuration allows us to capture the infall of low angular momentum gas blobs and molecular clouds that are predicted by chaotic cold accretion. Most of the CCA condensations occur around $10^4\rb$ (\newtext{$\rb \equiv GM/ c_{s,\infty}^2$ is the Bondi radius, discussed further in ~\ref{sec:bondi_test}} ) and start major collisions around $10 - 100\rb$, contributing to the AGN fueling process \citep{gaspari2017cca}. Capturing this process is crucial to studying the self-regulation of AGN feedback. One of the major challenges is to distinguish the outflow from eddies generated by the infalling clouds, as well as from the disk wind and the jet.

The accretion rate ${\dot M}$ is calculated by estimating the mass flux across a spherical control surface centered on the AGN and having a radius $\rcs$. This is more physically realistic than calculating the mass gain across a number of cells or just using the Cartesian flux returned from the Riemann solver. We use a Monte Carlo method, in which we choose $N$ random points on the surface with different angles ($\theta_i, \phi_i$), and sum the mass flow over these points to obtain
\begin{equation}
\dot{M} =  -\rcs^2 \int d \Omega\, \rho {\bf v} \cdot \hat{\bf n} 
\approx -\frac{4 \pi  \rcs^2}{N} \sum_{i=1}^N \rho_i v_{r,i}\ .
\end{equation}
Here $\rho_i$, ${\bf v}_i$, and ${\bf n}_i$ are the density, velocity, and unit outward normal vector for the $i$th control surface point. \newtext{To ensure the points are uniformly distributed on the sphere, we use uniform deviates $u$, $v$ and calculate ($\theta_i, \phi_i$) as $\theta_i$ = 2$\pi$u and $\phi_i$ = $\arccos$ (2v-1). }

Only inflow points (where $v_{r,i} = {\bf v}_i\cdot\hat{\bf n}_i \leq 0$) are counted, since we need to include all the material that has been falling onto the SMBH but exclude the jet. Two methods are considered: one is to count zero at points where there is an outflow, so that there are fewer than $N$ inflow points; another is to discard each outflow point and keep searching until we have $N$ inflow points. For our current jet configuration, we have run both methods and find the difference to be negligible since the jet only covers a small fraction of the control surface. However, since in the general case outflow can take place over a larger portion of the control surface, we use the second method in the calculations described here to minimize noise in the estimate of the integral.

\begin{figure}
    \centering
    \includegraphics*[width=1.\columnwidth]{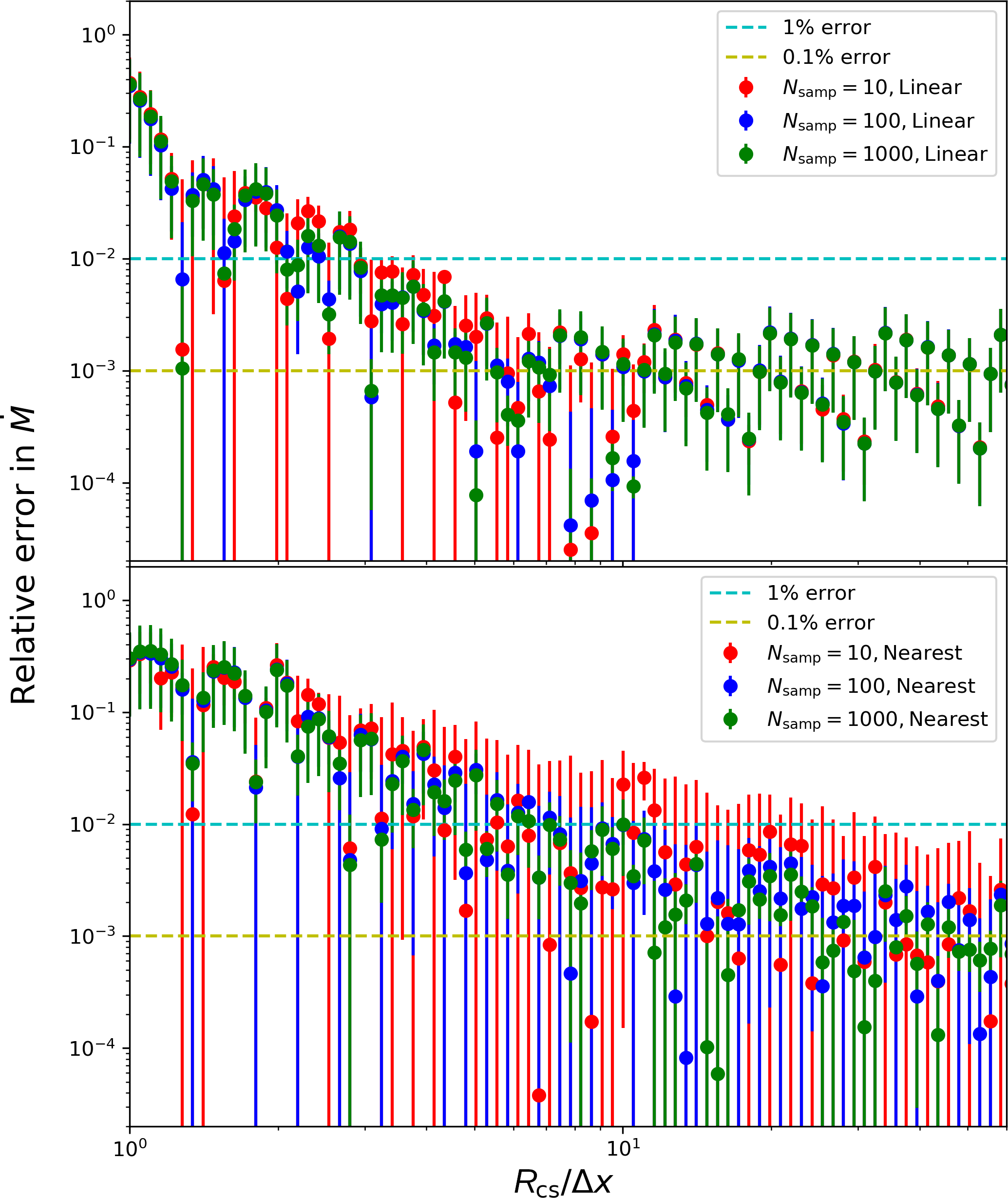}
    \caption{Error in the measured accretion rate (with respect to the analytical Bondi formula) and its relationship with the control surface radius in our accretion calculation, with different number of samples, and different in-grid interpolation techniques. The dashed lines represent 1\% and 0.1\% error respectively. }
    \label{acctest}
\end{figure}

We tested our flux calculation by initializing a Bondi profile setup (details of which are outlined in \S~\ref{sec:bondi_test}) on a uniform grid, and compared the results from the different flux methods to the analytical Bondi formula (Equation~\ref{eqn:bondi_rate}).
We estimated the uncertainty in this accretion rate calculation by varying two parameters: the number of random points, and the interpolation method from the mesh onto the control surface. For the latter, we used either tri-linear or nearest-neighbor interpolation. Figure~\ref{acctest} shows the relative error with respect to Equation~\ref{eqn:bondi_rate} as a function of control surface radius and number of samples $N=N_{\rm samp}$. Tri-linear interpolation gives smaller error for a given number of sample points and $\rcs$ value. The error scales with $\sqrt{N}$ to a certain radius ($\rcs \sim 250 \Delta x$, the maximum box size used), and the variation of error decreases with increasing number of sample points. We adopt $N = 1000$ and the tri-linear interpolation method for later calculations. The same approach can also be used to compute momentum and angular momentum fluxes when they are included in the SMBH configuration.

\subsubsection{Removing Gas from the Grid}
\label{sec:reset}

In our simulations we are not directly resolving the accretion disk, so after the accretion rate is calculated we must remove the gas accreted and effectively create an inner boundary that absorbs the inflow onto the black hole. Analogously to common sink particle algorithms in AMR simulations, this is handled by resetting the gas density and pressure within a spherical region of radius $\rdep$ around the black hole. The gas velocity is left unchanged, while the internal energy is reset using the equation of state.

To remove gas from the grid, we use a multiplicative resetting kernel whose value equals one at $\rdep$ to make the modified values continuous across the surface of the resetting region.
The resetting kernel is defined to be
\begin{equation}
\psi(r) = \psi_0 \left[\omega(r) - e^{-1}\right] + 1\ , 
\end{equation}
where
\begin{equation}
\omega(r) = \exp\left(-r^2/\rdep^2\right)\ . 
\end{equation}
Here $\psi_0$ is the normalization factor, which is calculated by requiring that the total gas removed be equal to the amount of gas accreted in a timestep, ${\dot M}\Delta t$. We set the new density in zone $ijk$ to be (with $n$ representing the current timestep and $n+1$ the next step)
\begin{equation}
\rho^{\rm n+1}_{ijk} = \rho^n_{ijk} \psi(r_{ijk})\ ,
\end{equation}
where $r_{ijk}$ is the radial position of the zone center. We require that the mass removed satisfy
\begin{equation}
-\dot{M} \Delta t = \int \psi \rho^n dV - \int \rho^n dV\ .
\end{equation}
Given the form of $\psi$, we have for the normalization factor
\begin{equation}
\psi_0 = \frac{\dot{M} \Delta t}{e^{-1} \int \rho^n\, dV - \int \omega \rho^n\, dV}\ ,
\end{equation}
so that
\begin{equation}
\label{eqn:newrho}
\rho^{n+1} = \rho^n \left(1+ \frac{\dot{M} \Delta t \left(\omega - e^{-1}\right)}{e^{-1} \int \rho^n\, dV - \int \omega \rho^n\, dV} \right)\ . 
\end{equation}
The volume integrals in these expressions are computed by summing the respective quantities within the cells inside the spherical reset region.

To ensure that no zones within the resetting region fall below a minimum density $\rho_{\rm min}$, we impose an additional constraint on the resetting kernel.
For each zone inside the reset radius, we require
\begin{equation}
   \rho^{n+1}_{ijk} = \rho^n_{ijk}[1-\psi_0(\omega_{ijk}-e^{-1})] \ge \rho_{\rm min}\ .
\end{equation}
Each of these zones thus has an allowed maximum value for the normalization factor $\psi_0$:
\begin{equation}
\psi_{0,{\rm max},ijk} = \frac{1-\rho_{\rm min}/\rho^n_{ijk}}{\omega_{ijk}-e^{-1}} \ .
\end{equation}
Hence our final choice for $\psi_0$ is
\begin{equation}
    \psi_{0,{\rm final}} = \min\left\lbrace \psi_0, \min_{r_{ijk}\le \rdep} \psi_{0,{\rm max},ijk}\right\rbrace\ .
\end{equation}
In our simulations, we have set $\rho_{\rm min}$ to be $10^{-33} {\rm\ g\ cm}^{-3}$ to avoid numerical instability. 

Similarly, we can reset the pressure according to the change of the internal energy density within the control surface per time step. We note however, since the transport equation for pressure has both advection and compression/expansion parts, simply measuring a flux for the pressure is not a correct approach. However, we can still use the change in pressure within the control region due to the hydrodynamic solver to determine how much internal energy is accreted. Assume that the change in energy density satisfies
\begin{equation}
-\dot{u} \Delta t = -\frac{1}{\gamma-1} \int (P^{*} - P^n)\, dV \ ,
\end{equation}
where $P^{*}$ is the updated pressure returned from the hydrodynamics equations and the physical processes that follow (e.g.\ radiative cooling), and $\gamma$ is the adiabatic index. \newtext{Here we have adopted $\gamma = 5/3$ for non-relativistic gas.} We thus have
\begin{equation}
\label{eqn:newp}
P^{n+1} = P^{*} \left(1+ \frac{(\gamma -1) \dot{u} \Delta t \left(\omega - e^{-1}\right)}{e^{-1} \int P^{*}\, dV - \int \omega P^{*}\, dV} \right)
\end{equation}
This will also be adapted in case we need to modify $\psi_0$ in order not to over-remove the mass accreted.

At the end of the timestep, we calculate the integrals by summarizing the gas property values in the region within $\rdep$, then calculate the fluxes and reset the gas properties. Then we calculate the new integral values that will be used to evaluate the fluxes returned from the physical processes in the next timestep. We leave the velocity components unchanged to avoid any numerical instabilities. This creates a non-reflecting boundary, and the gas properties inside the boundary do not impact the structure of the flow outside.

\subsection{Tests of the Accretion Model}

As verification tests, we run two sets of controlled experiments: the Bondi problem and the Bondi-Hoyle-Lyttleton (BHL) problem. The setup for each problem allows us to compare the simulation outputs to analytical results and published simulation work.

\subsubsection{Bondi Test}
\label{sec:bondi_test}

For this test, we assume that the accretion is spherically symmetric, and that a point mass $M$ is embedded in the ambient gas. This problem was first addressed in \cite{Bondi}. Far from the accreting object, the medium has a uniform density $\rho_{\infty}$ and a uniform pressure $P_{\infty}$. For adiabatic gas where $P \propto \rho^{\gamma}$, the sound speed far away should have the value $c_{s, \infty} = \sqrt{\gamma P_{\infty}/\rho_{\infty}}$. For isothermal gas, instead we have $c_{s,\infty} = \sqrt{ P_{\infty}/\rho_{\infty}}$.

We define the dimensionless variables
\begin{equation}
\begin{aligned}
x &\equiv \frac{r}{\rb},\ u \equiv \frac{\left| v \right|}{c_{s,\infty}}, \\
\alpha &\equiv \frac{\rho}{\rho_{\infty}}, \ \lambda \equiv \frac{\dot{M}}{4 \pi \rho_{\infty} (GM)^2 / c_{s,\infty}^3},
\end{aligned}
\end{equation}
where $\lambda$ is the dimensionless accretion rate, and $\rb$ is the Bondi radius $\rb = GM/c_{s, \infty}^2$. Then we can solve the mass and momentum conservation equations in a dimensionless form
\begin{equation}
u \alpha x^2 = \lambda
\end{equation}
\begin{equation}
\frac{u^2}{2} + H (\alpha) - \frac{1}{x} = 0
\end{equation}
where $H(\alpha)$ is given in polytropic flow by
\begin{equation}
H(\alpha) = \frac{1}{\gamma -1} (\alpha^{\gamma-1} - 1)\ \ \ \rm (\gamma \neq 1) 
\end{equation}
We use a separate program to solve the above set of differential equations and generate the Bondi solution for density, pressure and velocity profiles, then use that to initialize our grid. The theoretical Bondi accretion rate is
\begin{equation}
    \label{eqn:bondi_rate}
    \dot{M}_{\rm B} = 4 \pi \lambda \frac{G^2 M^2 \rho_{\infty}}{c_{s,\infty}^3}\ .
\end{equation}
For $\gamma = 5/3$, the parameter $\lambda = 0.25$.

We have conducted a number of runs with different accretion and reset radii, with the run parameters shown in Table~\ref{table1}. For convenience, $P_{\infty}$ and $\rho_{\infty}$ and $M$ values we use resemble the Perseus cluster ($P_{\infty} = 3.2 \times 10^{-10}\ {\rm dyn\ cm}^{-2}$, $\rho_{\infty} = 6.58 \times 10^{-26}\ {\rm g\ cm}^{-3}$), and we use a fixed $\gamma = 5/3$. The gravitational potential is set to be a fixed point mass potential with a central SMBH of mass $M = 3.0 \times 10^9\ M_{\odot}$. With this setup, the maximum free fall timescale is $t_{\rm ff, max} \sim \sqrt{L^3/GM} \sim 1.11 {\rm\ Myr}$.

\begin{table}
\begin{center}
  \caption{Table of runs conducted for the Bondi test. The last column describes what kind of resetting is done, where ``Bondi'' means resetting using the Bondi solution, and ``Kernel'' using our subgrid kernel method. }
  \begin{tabular}{l  c  c  c  r }
    \hline
     Run Number & $\Delta x / \rb$ & $\rcs/\rb$ & $\rdep/\rb$ & Method \\ \hline
     W0 & 1/16 & 1 & N/A & None\\ 
     W1 & 1/16 & 1 & 1 & Kernel\\ 
     W1B & 1/16 & 1 & 1 & Bondi\\ 
     W2 & 1/16 & 1 & 1/2 & Kernel \\ 
     W3 & 1/16 & 1/2 & 1/2 & Kernel \\
     P0 & 1/4 & 1 & N/A & None \\
     P1 & 1/4 & 1 & 1 & Kernel\\ 
     P1B & 1/4 & 1 & 1 & Bondi \\
    \hline
  \end{tabular}
  \label{table1}
\end{center}
\end{table}

\begin{figure*}
    \begin{center}
    \includegraphics*[width=1.\columnwidth]{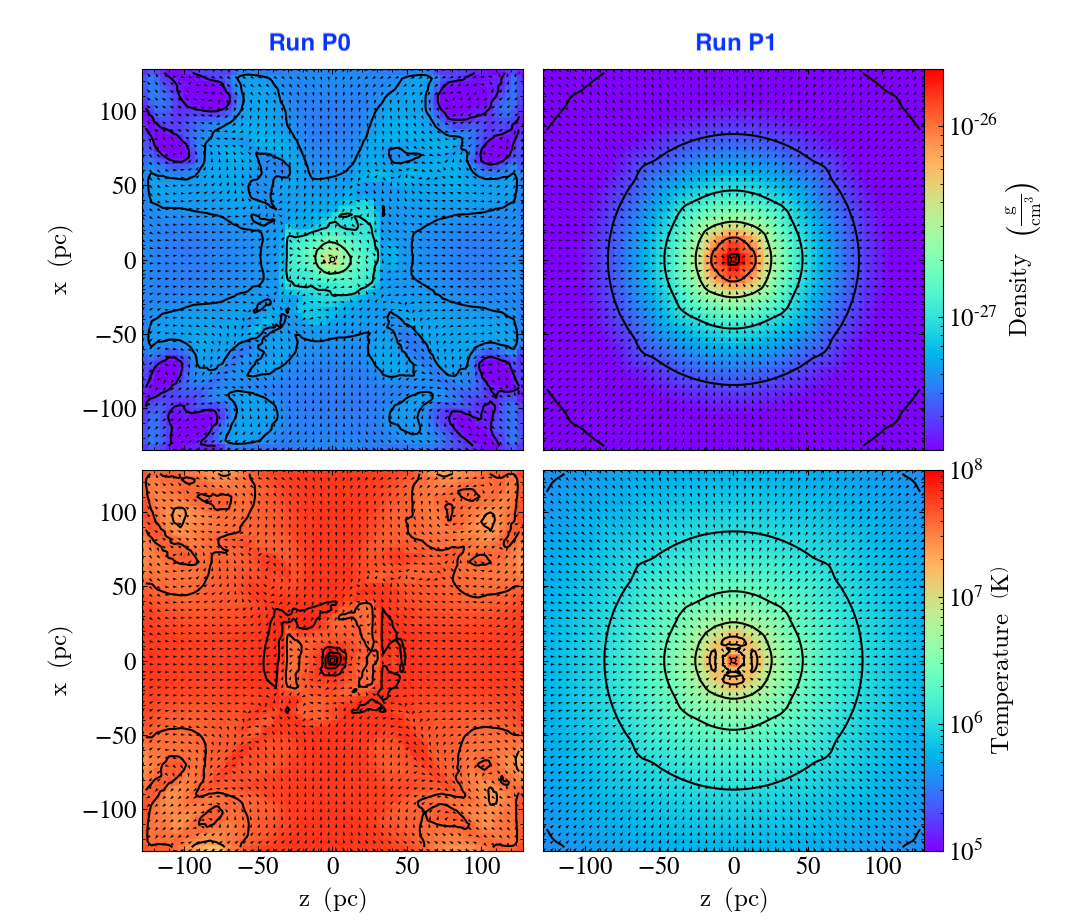}
    \includegraphics*[width=1.\columnwidth]{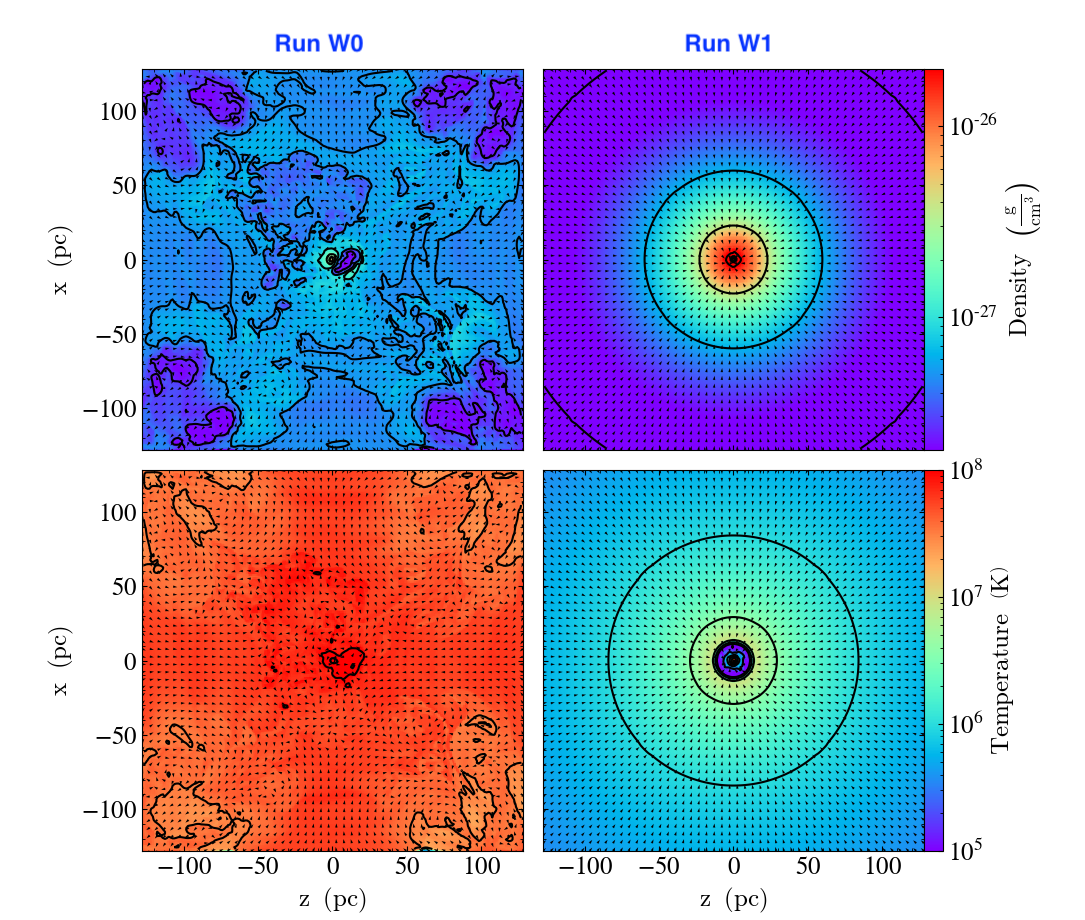}
    \caption{Left: Density and temperature slice plots at $t = 1$~Myr for runs without reset (left) and with kernel reset (right) in the poorly-resolved case (P runs). Right: Density and temperature slice plots at $t = 1$~Myr for runs without reset (left) and with kernel reset (right) for the well-resolved case (W runs).}
    \label{slice_bondi}
    \end{center}
\end{figure*}

Figure~\ref{slice_bondi} depicts the flow for runs with different resolutions. In both well-resolved and poorly-resolved cases, without resetting the gas there is an accumulation of material around the center, eventually developing a back pressure that influences the flow of the gas outside the control surface and leads to an increase in the density and temperature there. However, by resetting the gas we are able to maintain a steady inflow, as is predicted by the theoretical Bondi solution. Moreover, while the kernel reset method knows nothing about the ``correct'' solution, it reduces the back pressure and converges to the Bondi-resetting result outside the control surface.

\begin{figure}
    \centering
    \includegraphics*[width=1.\columnwidth]{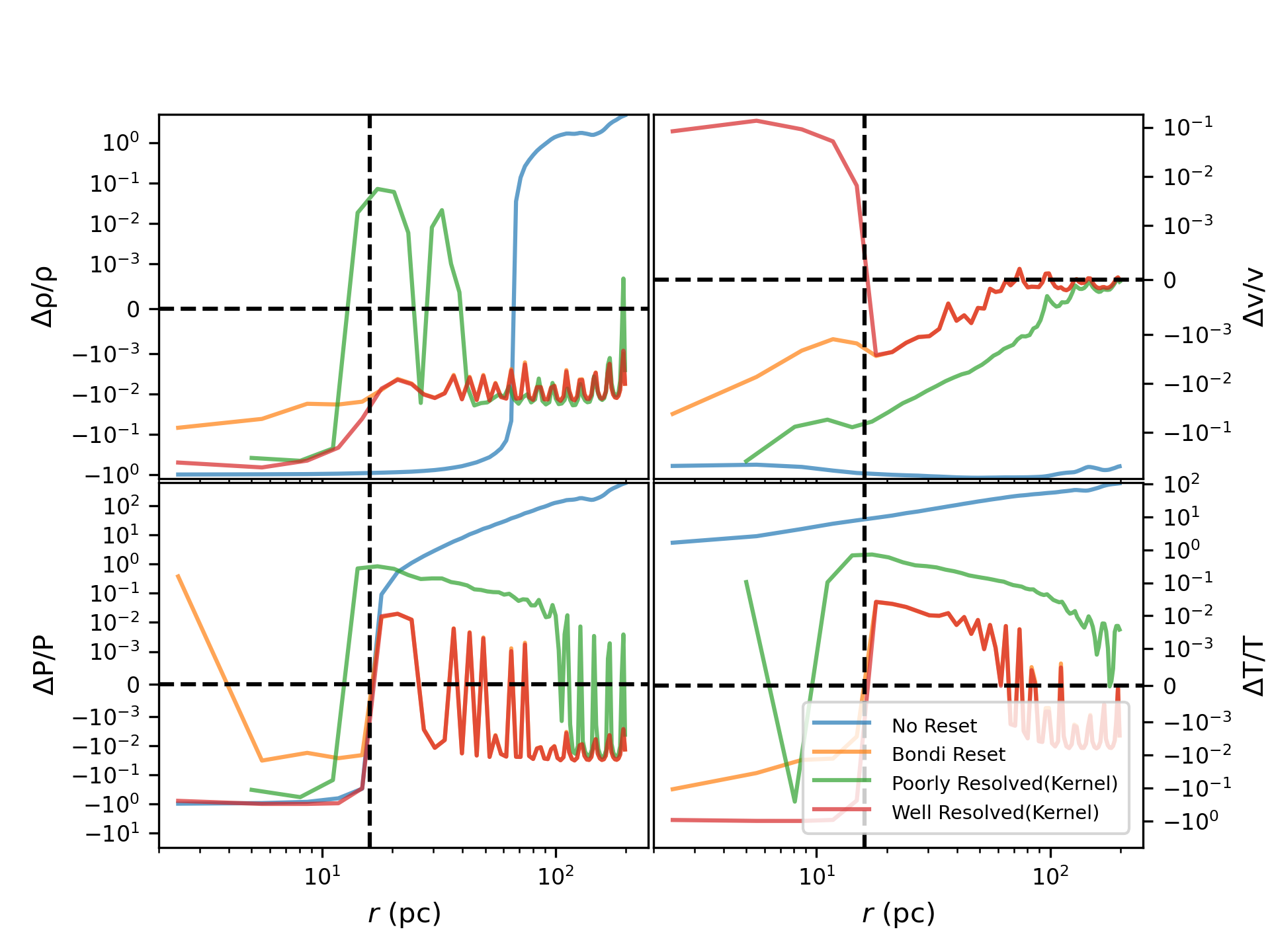}
    \caption{Density, pressure, temperature, and velocity residual profiles of the Bondi test runs, with and without resetting, in comparison with the Bondi solution, taken at $t = 1$~Myr, around 0.9 of the free fall timescale. Plotted here are runs W0 (blue solid line), P1 (green solid line), W1B (orange solid line) and W1 (red solid line). The temperature calculation assumes a 75\% hydrogen/25\% helium composition. The over-plotted vertical line is the Bondi radius, and the horizontal line represents the ambient value.} 
    \label{prof_bondi}
\end{figure}

This can be demonstrated quantitatively using the radial gas profiles, measured from the center. In Figure~\ref{prof_bondi} we show the deviation of these properties from the Bondi solution, since the difference in absolute values of various gas properties is very small. Through the profiles we can further see the influence of gas properties with and without resetting. With kernel resetting, although we see larger errors inside the control surface compared to resetting directly to the Bondi solution, for flow outside $\rcs$ the difference is negligible (within $10^{-3}$ error). Higher resolution improves the agreement with the Bondi solution especially around the Bondi radius, most significantly in the flow velocity and pressure. 

\begin{figure}
\begin{center}
    \includegraphics*[width=.9\columnwidth]{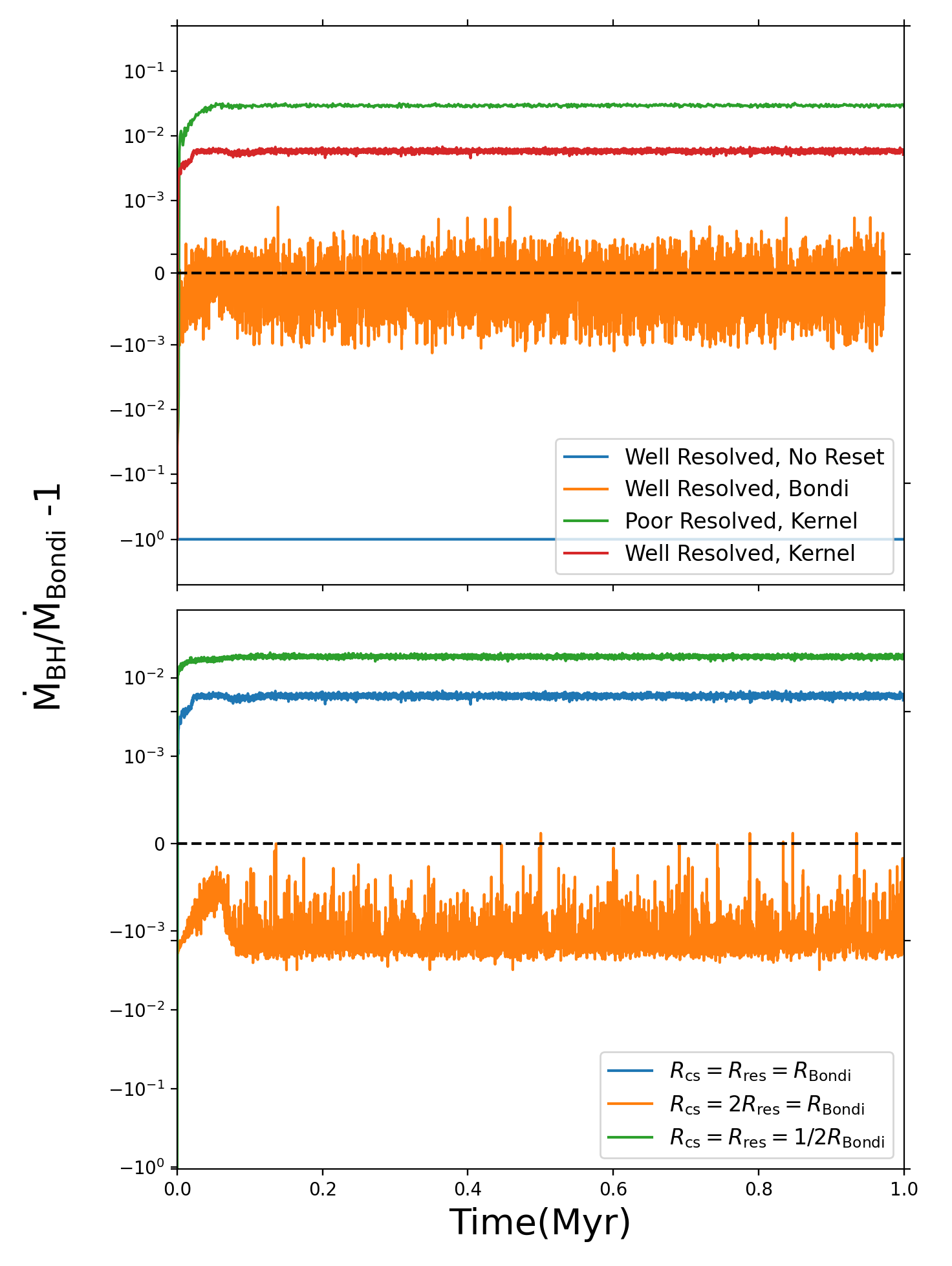}
    \caption{Accretion rate and comparison to the predicted Bondi rate for different Bondi test runs plotted as $\dot{M}/\dot{M}_{\rm B} -1$. This value equals zero when the measured accretion rate is exactly the same as the predicted value. Top panel: Accretion rate with different resetting methods. Plotted here are runs W0 (blue solid line), P1 (green solid line), W1B (orange solid line) and W1 (red solid line). Bottom panel: resetting with a kernel reset for a well-resolved case, but at different accretion and reset radii. Plotted here are W1 (blue), W2 (orange) and W3 (green).  }
\label{macc_bondi}
\end{center}
\end{figure}

This result is further supported by the accretion rate, as shown in Figure~\ref{macc_bondi}. With resetting the accretion rate remains close to the Bondi accretion rate predicted by the analytical formula. The most accurate measurement of accretion rate comes from defining the control surface radius to be the same as the Bondi radius. If keeping the control surface fixed to be the Bondi radius, decreasing the resetting radius can decrease the error of the accretion rate calculation. In general, the best outcome comes from using the same value for the control surface radius and the reset radius (both equaling the Bondi radius).

In comparison with a similar published test result \citep{Beckmann}, our method is able to better reproduce the accretion rate; we find $|\dot{M}/\dot{M}_{\rm B} -1| < 0.01$ for the well-resolved setup ($\rb/\Delta x=16$) as compared to their result of $|\dot{M}/\dot{M}_{\rm B}-1| \sim 0.22$ for the same resolution. We do observe a similar convergence behavior where the gas properties outside the Bondi radius vary with resolution: under-resolved runs tend to underestimate the flow velocity and overestimate pressure and temperature in the few zones around the Bondi radius.   

\subsubsection{Bondi-Hoyle-Lyttleton (BHL) Test}

The Bondi problem represents the low Mach number limit of the problem of accretion onto a point mass from a moving uniform medium. In the high Mach number limit, the accretor moves at a speed $v_\infty \gg c_{s,\infty}$ with respect to the medium, and the accretion rate becomes \citep{Hoyle}
\begin{equation}
    {\dot M}_{\rm HL} = \pi \rhl^2\rho_\infty v_\infty\ ,
\end{equation}
where
\begin{equation}
    \rhl = 2 \frac{GM}{v_\infty^2}\ .
\end{equation}
For the general case an interpolation formula must be used for the accretion rate; one that is accurate to within 10\% or so \citep[see][and references therein]{Ruffert} is
\begin{equation}
   \label{eqn_bhl}
   {\dot M}_{\rm BHL} = 4\pi \rbhl^2\rho_\infty(\lambda^2 c_{s,\infty}^2+v_\infty^2)^{1/2}\ ,
\end{equation}
where
\begin{equation}
    \rbhl = \frac{GM}{c_{s,\infty}^2 + v_\infty^2}\ .
\end{equation}

This is named as the Bondi-Hoyle-Lyttleton problem, and for this test, we use the above result to test the performance of our subgrid model for moving flows. The setup is similar to the Bondi test, but we apply a 45$^\circ$ angled inflow boundary condition on two sides of the box and a diode boundary condition on the other sides. While this problem does not have an analytic solution, it has been studied extensively via analytic approximations and numerical simulations; see \cite{Edgar} for a review.

To compare with previous work, we ran the test with various black hole masses and Mach numbers ${\cal M}\equiv v_\infty/c_\infty$, to demonstrate resolutions for which $\rbhl$ is both well-resolved and under-resolved. We also ran tests with different control surface and reset radii, with the full range of run parameters shown in Table~\ref{table_bhl}. However, to ensure enough information is contained on the control surface, we set the minimum control surface and reset radii to be at least two zones across. This means that for runs where the BHL radius is unresolved, resetting is still done, but over a larger region than the gravitational capture radius. The runs are executed on a uniform grid with resolution $\Delta x = 0.25{\rm\ pc}$ in a 3D box with size $L_{\rm box} = 128{\rm\ pc}$ in each dimension. To encompass a range of resolutions, we set the fiducial black hole mass to be $M = 1.0 \times 10^9\ M_{\odot}$ and the ambient density to be 1/3 of the value used in the Bondi test, while other initial setup parameters remain the same. 

From previous studies, one should expect a bow shock to form in front of the SMBH, raising the temperature of the gas and decreasing its bulk velocity. At radii $r \ll \rbhl$, the flow of the gas is approximately radial, and takes the form of a spherically symmetric Bondi solution. 

Since the accretion radius changes as the black hole grows, we considered a different approach to setting the control surface radius. This is the ``flexible'' option in run A1. For the flexible run, we re-evaluated the wind velocity by averaging the gas properties within the Bondi radius of the accretor (which is an upper limit to the BHL radius, and thus large enough to incorporate the flow pattern around the accretor), then updated $\rcs$ to correspond to the new values. The new value is further constrained by having the control surface radius remain between $2\Delta x$ and $8\Delta x$, in order to avoid the development of numerical instability.  \newtext{For this problem, the} flexible method does not seem to impact the flow \newtext{measurably} compared to run A3. However, \newtext{in other scenarios} where the black hole grows significantly, the flexible option is useful to consider. 

%\begin{figure}
%    \begin{center}
%     \includegraphics*[width=1.\columnwidth]{bhl_slice_a1f_new.png}
%    \caption{Density slices from the BHL test runs with $\mathcal{M} = 3$, \newtext{plotted within a region of size 64}~pc, for fixed (A3) and flexible (A1) control surface radius. Black circle represents the accretion radius $\rbhl$.}
%    \label{bhl_slice0}
%    \end{center}
%\end{figure} 

\begin{table*}
\movetableright -0.5in
\begin{center}
    \caption{Table of runs conducted for the Bondi-Hoyle-Lyttleton test. Each run used the kernel resetting method with the reset radius and control surface radius indicated, scaled to the resolution of the run.}
    \begin{tabular}{ll c c c c r} 
        \hline
         Objective &
         Run Number &  $M/M_{\odot}$ & Mach Number & $\rbhl/\Delta x$ & $\rcs/\Delta x$ & $\rdep/ \Delta x$\\ \hline
         Vary control surface &A1 & 1.0 $\times 10^9$ & 3 & 6.36 & Flexible &  2\\   % 
         & A2 & 1.0 $\times 10^9$ & 3 & 6.36 & 16 &  2\\
         & A3 & 1.0 $\times 10^9$ & 3 & 6.36 & 8 &  2\\
         & A4 & 1.0 $\times 10^9$ & 3 & 6.36 & 4 &  2\\
         & A5 & 1.0 $\times 10^9$ & 3 & 6.36 & 2 &  2\\
        \hline
        Vary reset radius & B1 & 1.0 $\times 10^9$ & 3 & 6.36 & 8 &  8\\
        & B2 & 1.0 $\times 10^9$ & 3 & 6.36 & 8 &  4\\
        & B3 (Same as A3) & 1.0 $\times 10^9$ & 3 & 6.36 & 8 &  2\\
        & B4 & 1.0 $\times 10^9$ & 3 & 6.36 & 8 &  1\\
        \hline
        Vary accretion radius & C1 & 1.0 $\times 10^7$ & 3 & 0.06 & 2 &  2\\
        & C2 & 1.0 $\times 10^8$ & 3 & 0.64 & 2 &  2\\
        & C3 & 3.0 $\times 10^8$ & 3 & 1.91 & 2 &  2\\
        & C4 (Same as A5) & 1.0 $\times 10^9$ & 3 & 6.36 & 2 &  2\\
        & C5 & 3.0 $\times 10^9$ & 3 & 19.07 & 2 &  2\\
        & C6 & 1.0 $\times 10^{10}$ & 3 & 63.58 & 2 &  2\\
        \hline
        Vary Mach number & D1 & 1.0 $\times 10^9$ & 1 & 31.79 & 2 &  2\\
        & D2 (Same as A5)  & 1.0 $\times 10^9$ & 3 & 6.36 & 2 &  2\\
        & D3 & 1.0 $\times 10^9$ & 5 & 2.45 & 2 &  2\\
        & D4 & 1.0 $\times 10^9$ & 10 & 0.63 & 2 &  2\\
         \hline \hline
    \end{tabular}
        \label{table_bhl}
\end{center}
\end{table*}

\begin{figure*}
    \begin{center}
     \includegraphics*[width=1.\columnwidth]{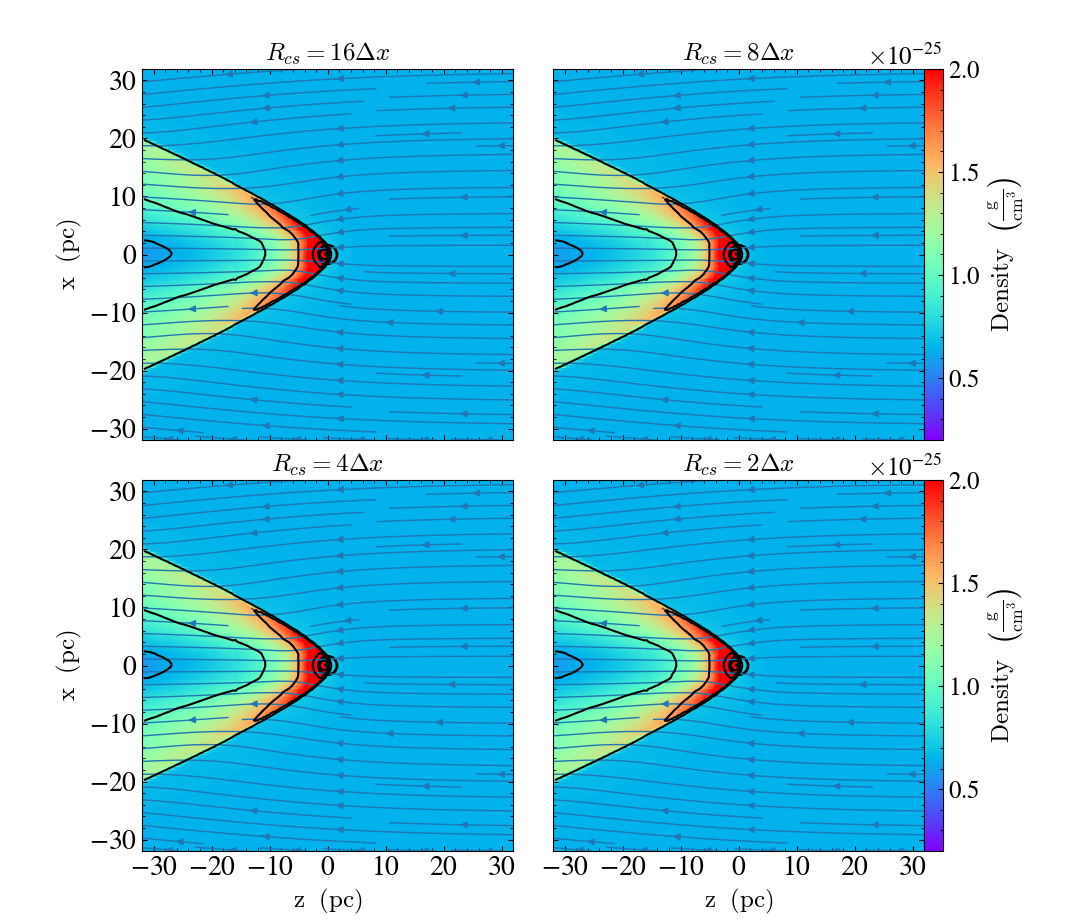}
     \includegraphics*[width=1.\columnwidth]{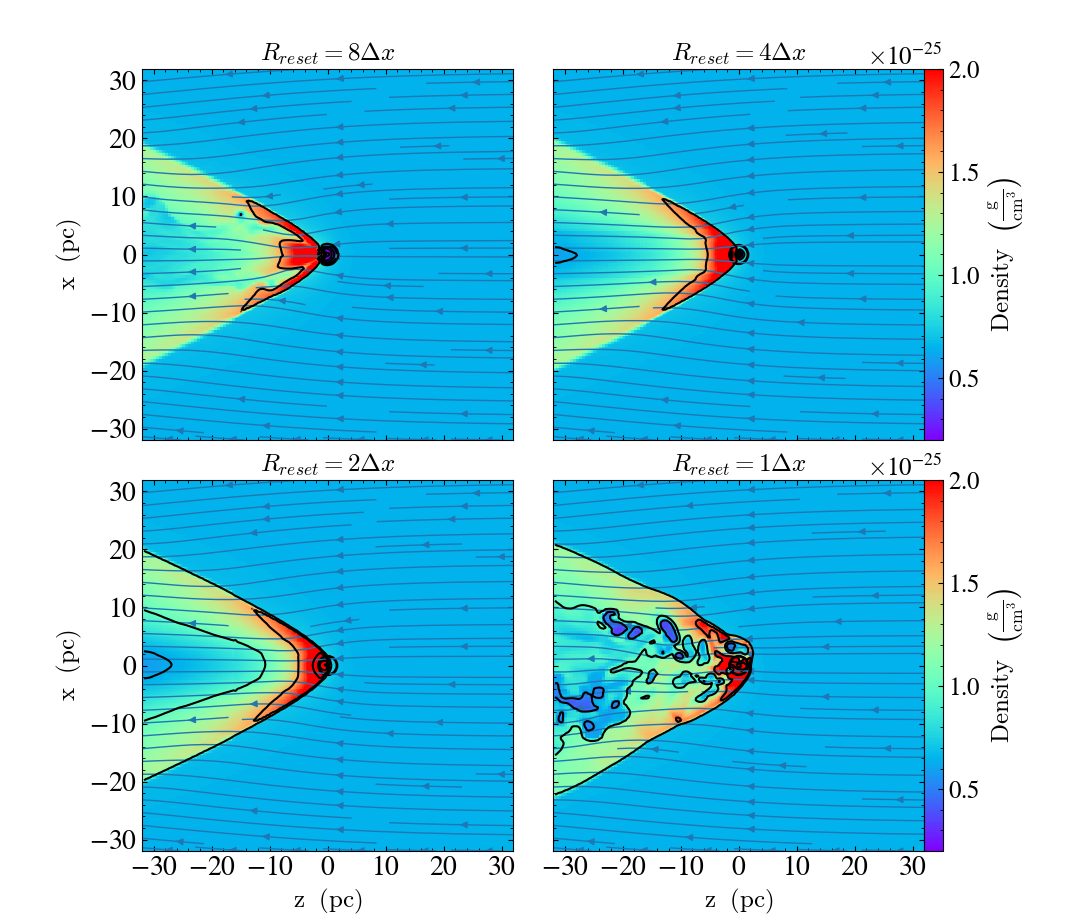}
    \caption{Density slices of the runs with $M = 10^9 M_{\odot}$ and $\mathcal{M} = 3$ at $t = 1 {\rm\ Myr}$ (run A1 to B4), plotted within a region of size $64\ {\rm pc}$. The left panel shows the A series, and the right panel shows the B series. Overplotted are density contours and velocity streamlines. Black circles represent the accretion radius $\rbhl$.}
    \label{bhl_slice1}
    \end{center}
\end{figure*}

\begin{figure*}
    \begin{center}
     \includegraphics*[width=2.\columnwidth]{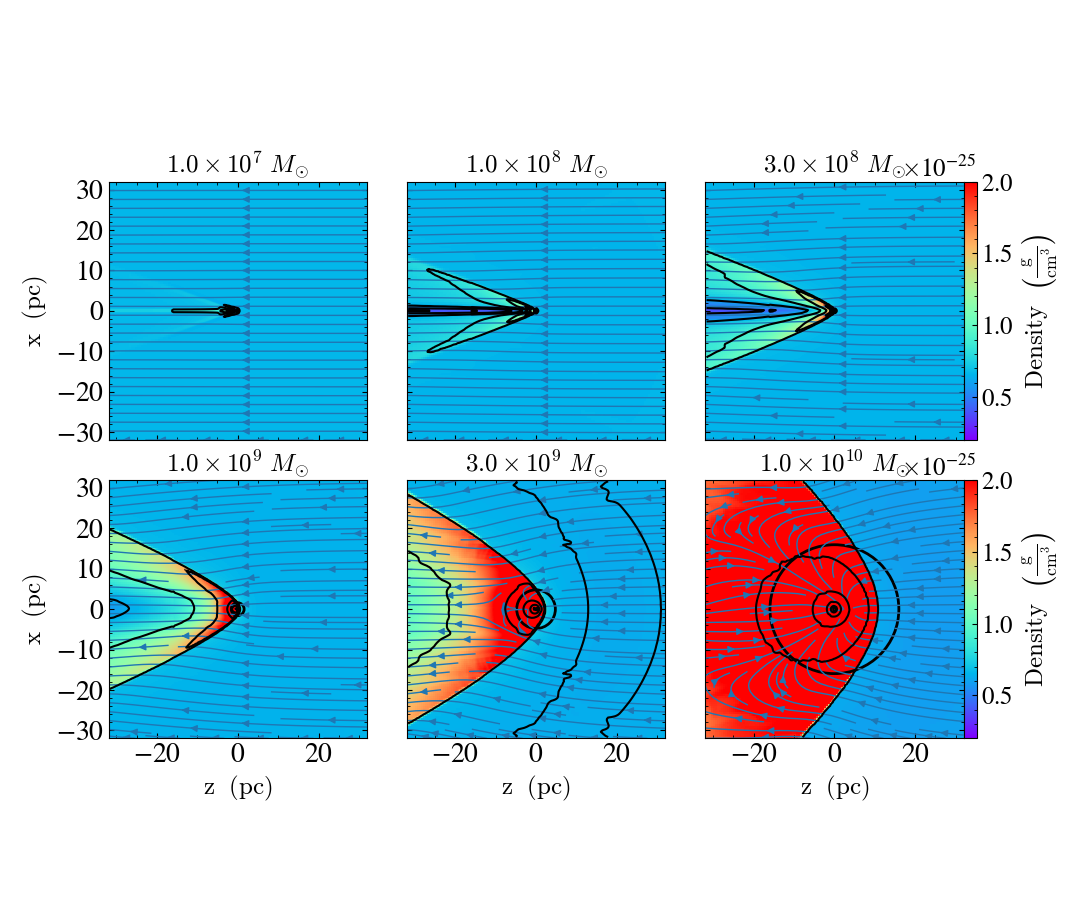}
    \caption{Density slices of the runs with different central black hole masses at $t = 1 {\rm\ Myr}$ (runs C1 to C6), plotted within a region of size $64\ {\rm pc}$. Black circles represent the accretion radius $\rbhl$.}
    \label{bhl_slice2}
    \end{center}
\end{figure*}

\begin{figure}
    \begin{center}
     \includegraphics*[width=1.\columnwidth]{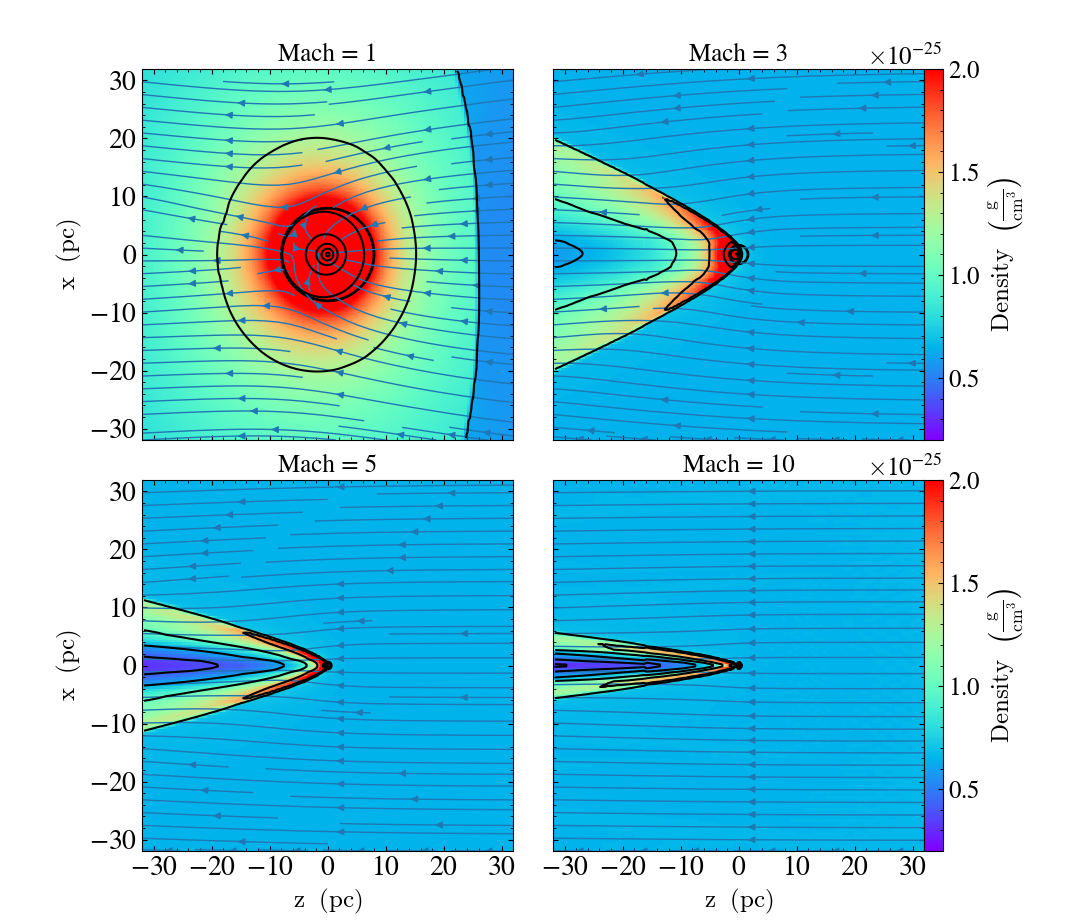}
    \caption{Density slices of the runs with different Mach numbers at $t = 1{\rm\ Myr}$ (runs D1 to D4), plotted within a region of size $64\ {\rm pc}$. Black circles represent the accretion radius $\rbhl$.}
    \label{bhl_slice3}
    \end{center}
\end{figure}

For the runs listed in Table~\ref{table_bhl}, we examine the flow pattern and gas dynamics, and we compare the measured accretion rate to Equation~(\ref{eqn_bhl}). In Figures \ref{bhl_slice1}, \ref{bhl_slice2}, and \ref{bhl_slice3} we show density slices with flow fields perpendicular to the $y$ axis at time $t = 1 {\rm\ Myr}$, which is the duration of the run. The flow has mostly reached steady state at this time for all our runs. 

\begin{figure}
      \includegraphics*[width=1.\columnwidth]{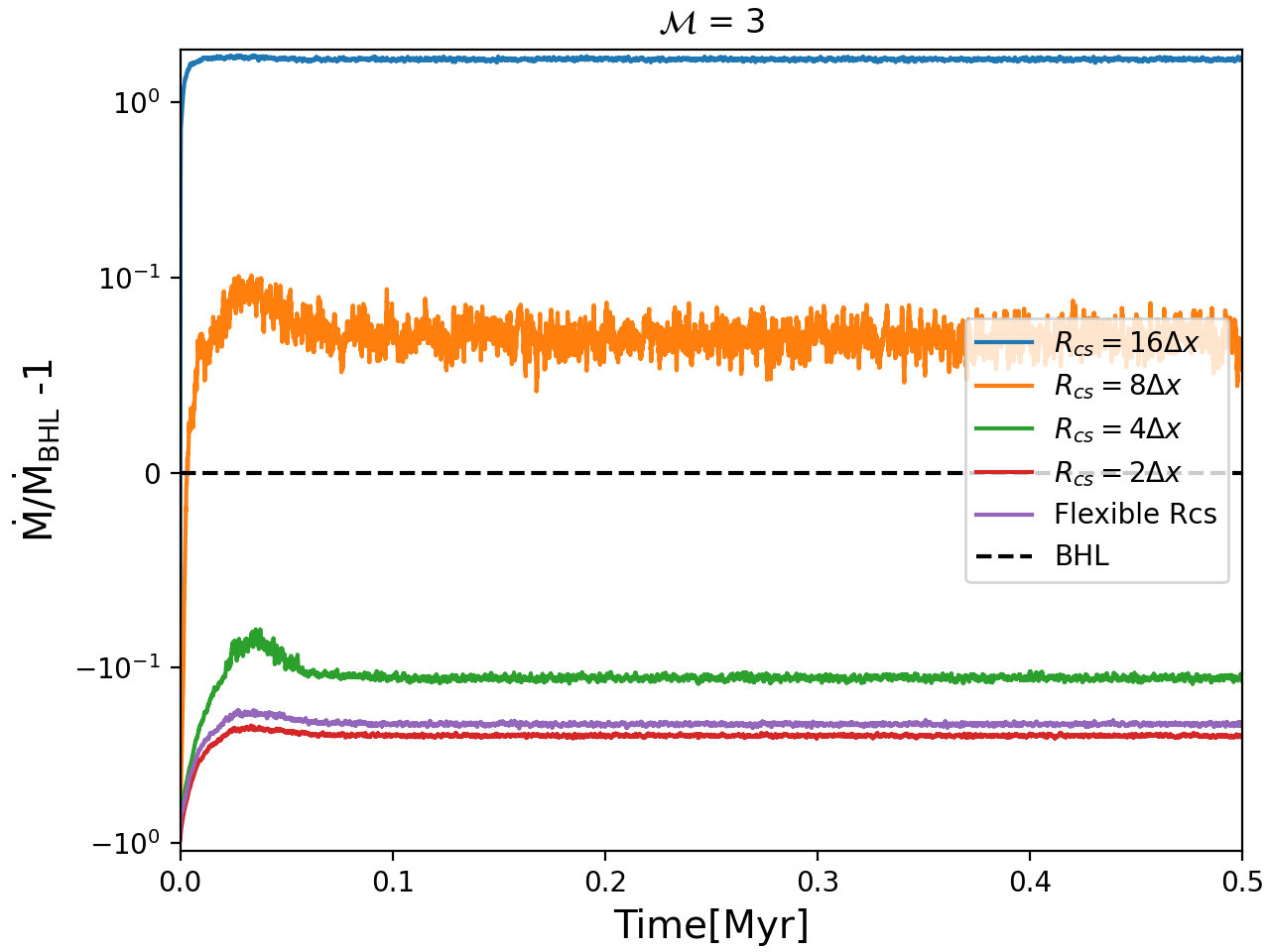}
      \includegraphics*[width=1.\columnwidth]{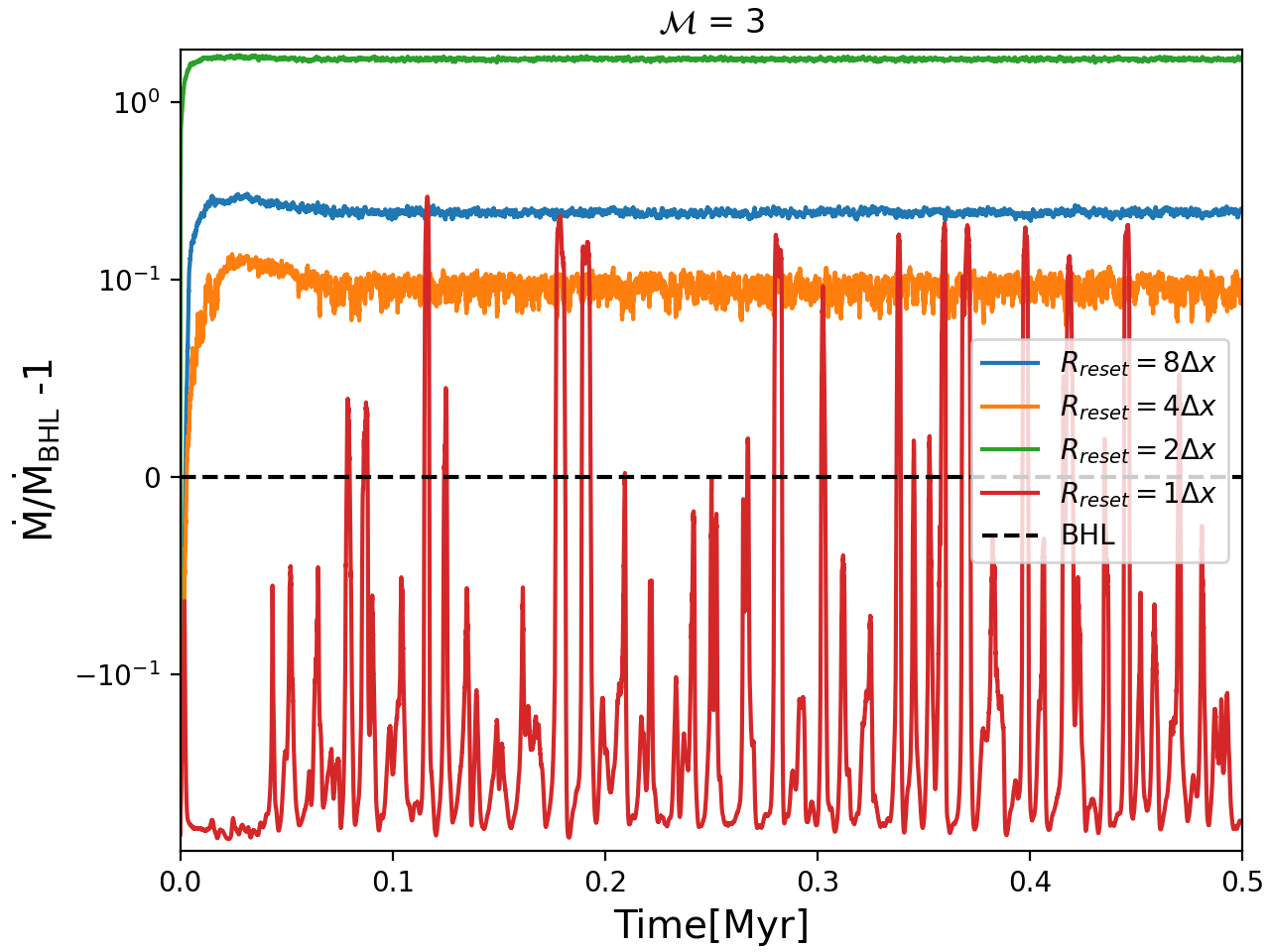}
    \caption{Accretion rate and comparison to the predicted Bondi-Hoyle-Lyttleton rates of different test runs, for A series (top) and B series (bottom). }
    \label{bhl_macc1}
\end{figure}

Figures~\ref{bhl_slice1} and \ref{bhl_macc1} show the gas flow behavior and accretion rate measured through the control surface for the A and B run series (fixing $\rbhl = 6.36\Delta x$), varying the control and reset radii. Not surprisingly, varying $\rcs$ does not affect the flow structure, since there is no feedback, but the measured accretion rate is different. Using a flexible control surface radius, the run shows that as the steady state solution develops, the accretion radius stays between 2 and $4 \Delta x$. The accretion rates are plotted as residuals from the BHL rate (Equation~\ref{eqn_bhl}). For smaller control surface radii, the measured accretion rate becomes more accurate. For fixed $\rcs$, as the reset radius decreases to $2 \Delta x$, the accretion rate converges to the BHL value and the shock moves from behind the accretor (unphysical) to in front of the accretor. However, reducing $\rdep$ to $1 \Delta x$ causes instability to develop behind the accretor that impacts the flow, also making what is measured through the control surface inaccurate. \newtext{The $\rdep = 2\Delta x$ case places the bow shock ahead of the resetting region, but produces a larger error in ${\dot M}$, while the $4\Delta x$ case places the shock within the resetting region and produces the most accurate accretion rate. To more easily explore the effects of resolution, we adopt} $\rcs = 2 \Delta x$ and $\rdep = 2 \Delta x$ for the remaining runs in this section. \newtext{However,} in the jet feedback tests described in \S~\ref{sec:jet_tests}, we use $\rcs = 8\Delta x$ and $\rdep = 4\Delta x$. 

\begin{figure}
      \includegraphics*[width=1.\columnwidth]{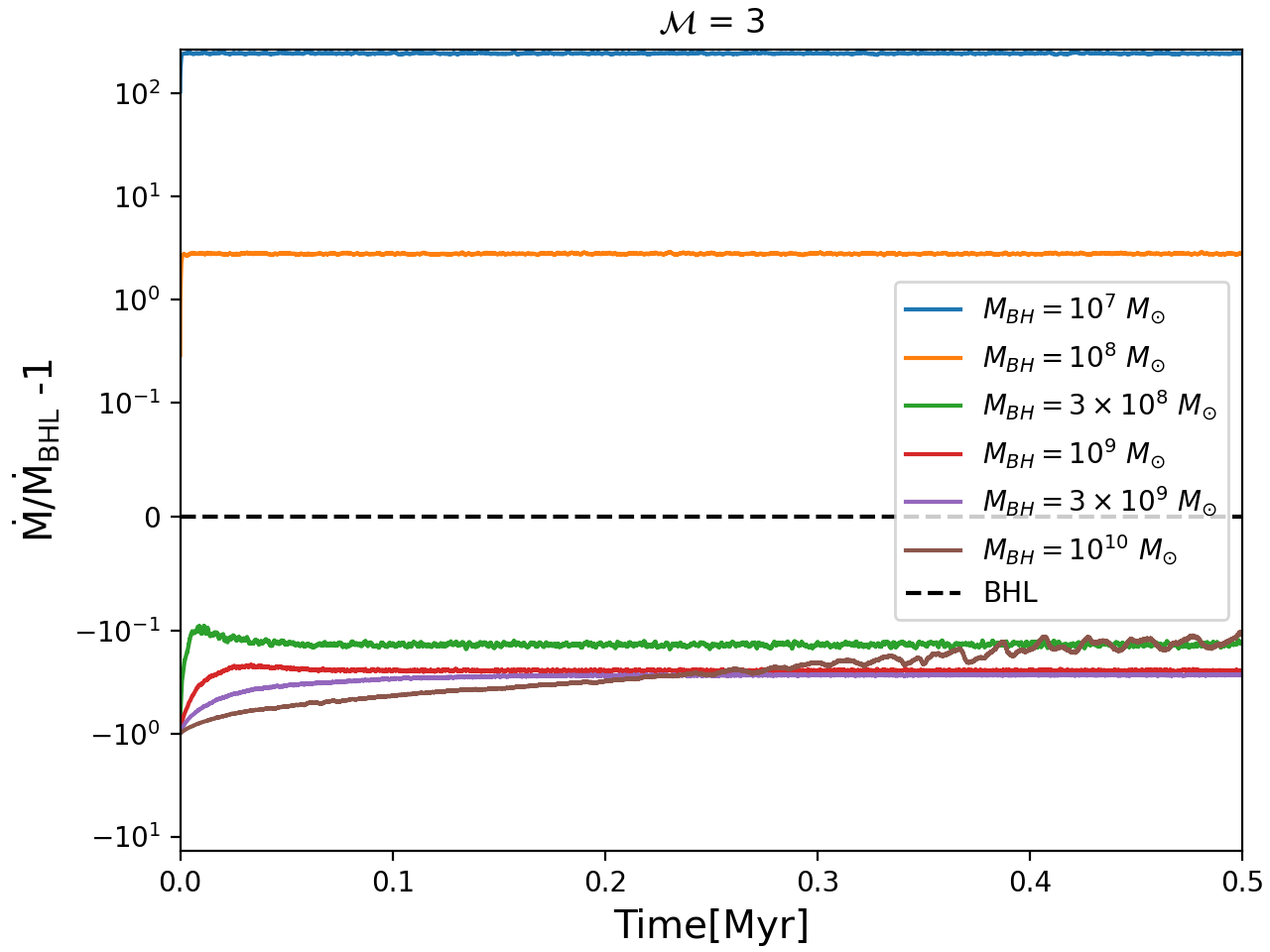}
      \includegraphics*[width=1.\columnwidth]{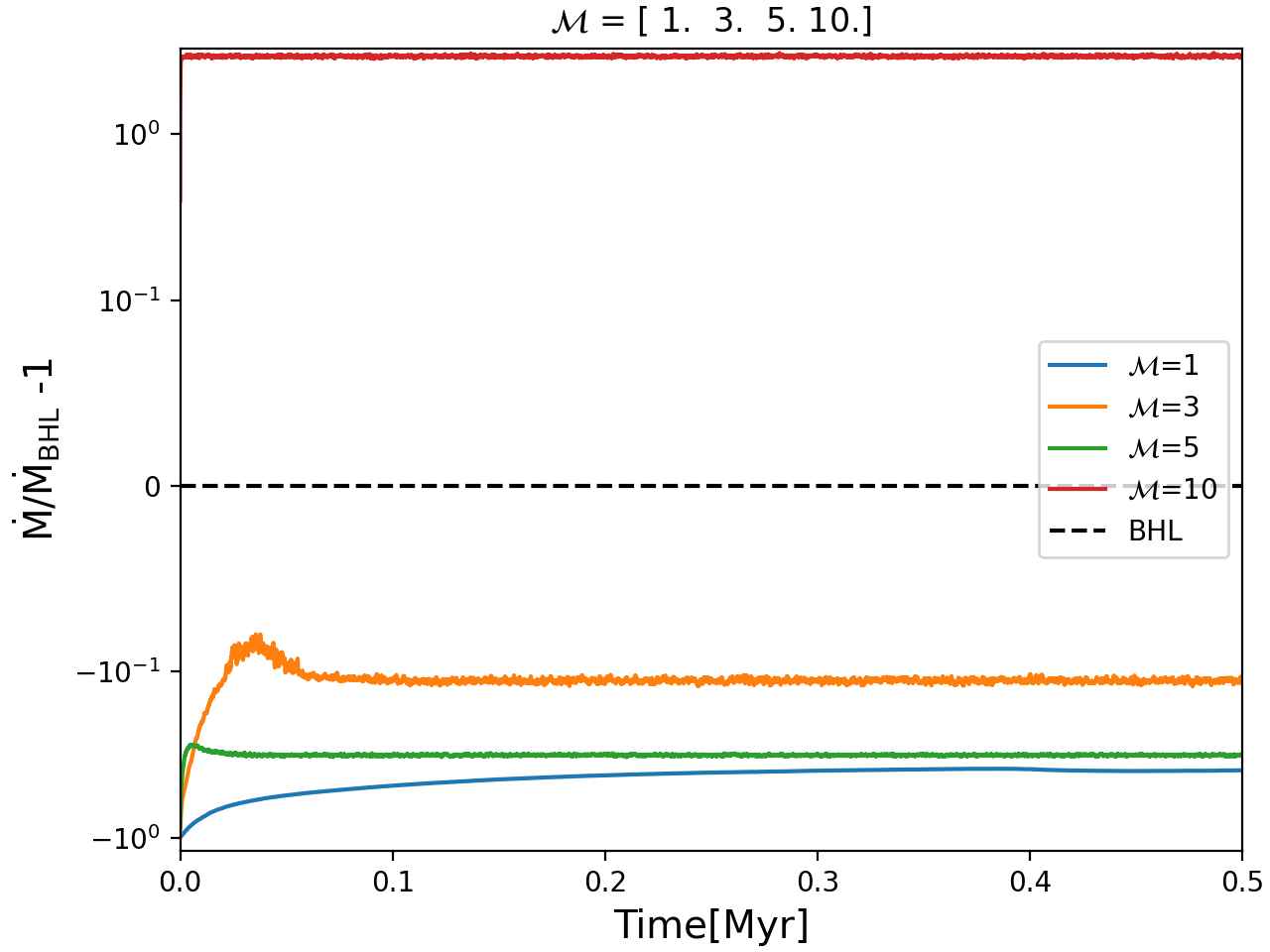}
    \caption{Accretion rate and comparison to the predicted Bondi-Hoyle-Lyttleton rates of different test runs, for C series (top) and D series (bottom). }
    \label{bhl_macc2}
\end{figure}

For the C series runs, we fix $\rcs$ and $\rdep$ but vary the mass of the black hole, thus changing $\rbhl$ from being under-resolved (low mass runs) to well-resolved (high mass runs). The expected behavior of the flow, where the shock forms in front of the accretor, only occurs with well-resolved runs, based on the slice plots shown in Figure~\ref{bhl_slice2}. Higher resolution leads to a better defined shock, while poorly resolved runs resemble an unperturbed flow more as the wind directly passes the accretor. For the higher mass runs, we measure a standoff distance for the bow shock (the distance between the stagnation point and the closest point on the shock front) to be around $0.2-0.3 \rbhl$, which is consistent with what Ruffert et al.\ has observed. Similarly, from the top panel of Figure~\ref{bhl_macc2} we observe a more accurate measurement of accretion rate for the well-resolved runs. So long as the control surface and reset radius are similar to or smaller than the BHL radius, the accretion and reset method can capture the flow successfully and reproduce the Bondi-Hoyle-Lyttleton solution to a large extent. 

Run series D, as shown in Figure~\ref{bhl_slice3} and the bottom panel of Figure~\ref{bhl_macc2}, reveals the flow structure for different Mach numbers ${\cal M}$. The opening angle of the shock is larger for smaller Mach numbers (with $\theta \sim 20^\circ$ for $\mathcal{M} = 5$ and $ \theta \sim 40^\circ$ for $\mathcal{M} = 3$), which is consistent with what has been observed in previous work. The shock stand-off distance also decreases with Mach number, consistent with  previous work such as \cite{Ruffert} and \cite{Verigin}. This behavior is also consistent with the C series given that higher Mach number makes $\rbhl$ much smaller and thus under-resolved in our runs. The convergence behavior with increasing resolution varies for different Mach numbers, however. The best estimate of accretion rate occurs for $\mathcal{M} = 3$, although as seen in the C runs, the error for other runs stays within 10\%, possibly due to greater back flow in our simulations. This corresponds to a bow shock jump higher than the other C runs.

In general, we show a BHL test with flow solution similar to what has been observed in most previous work, such as \cite{Ruffert} and \cite{Beckmann}.  The differences arise from the bow shock structure, as our resetting routine is different from what other simulators have used. Accretion rates measured with our method closely match the analytical BHL value, and compared with \cite{Beckmann}, the dependence of accretion rate on resolution shows a similar behavior. However, we are able to reproduce the expected accretion rate better than their results with similar Mach numbers and resolution configurations.  

In many previous works, Bondi-Hoyle-Lyttleton accretion can cause flip-flop instability in the flow inside the shock cone behind the accretor \citep{Taam, Foglizzo}. The shock surface oscillates from one side of the accretor to the other, causing the accretion rate to vary in a quasi-periodic behavior. This behavior is observed to be stronger when the accretor is smaller and the flow is more supersonic according to their work. In our test we do not observe a strong flip-flop instability, but we do see some evidence of instability for higher Mach number and more well-resolved runs.

\section{Jet Feedback Method and Tests}

\subsection{Modeling of the Jet}

Most previous grid-based AGN modeling efforts have implemented feedback, in the form of either jets or bubbles, by adding mass, momentum, and energy source terms to a fixed volume in the vicinity of the accretor \citep[e.g.,][]{Yang, Martizzi}. Since we are effectively treating the control surface as an embedded boundary, a more natural approach would seem to be to apply an outward flux on the control surface. A flux-based injection approach is used in dedicated jet simulations and has been used by \cite{bruggen07} in the context of AGN feedback. However, in general we must allow for arbitrary numbers of AGN particles with arbitrary spin axes. Since our grid is Cartesian, the spherical control surface does not correspond to a coordinate boundary, and the jet generally emerges at an angle to the grid. The control surface is also usually not well-resolved. Therefore, we have chosen to impose feedback in the traditional way, by adding source terms to zones within a cylindrical volume.

In this and the following sections we use $M_{\rm BH}$ instead of $M$ to refer to the black hole mass.

Following \cite{Yang.etal2012}, to impose jet feedback, the mass, momentum, and thermal energy injection rates per unit volume in a given zone are given by
\begin{eqnarray}
\dot{\rho} &=& \eta \dot{M}_{\rm BH} |\Psi|  \\
\dot{p} &=& \sqrt{2\eta\epsilon_{\rm f}(1-\epsilon_{\rm m})} \dot{M}_{\rm BH} c \Psi \\
\dot{E}_{\rm th} &=& \epsilon_{\rm m}\epsilon_{\rm f} \dot{M}_{\rm BH} c^2 |\Psi| \ ,
\end{eqnarray}
where $\epsilon_{\rm f}$ is the feedback efficiency, $\epsilon_{\rm m}$ is the thermal energy factor,
$\eta$ is the mass loading factor, and $\Psi$ is a function of position that describes the jet injection region. For fiducial values, we use $\epsilon_{\rm f} = 0.0625$, $\epsilon_{\rm m} = 0.8$, and $\eta = 2$, making the jet velocity $v_{\rm jet} = 0.112 c$, which is mildly relativistic. The jet efficiency value is similar to common values used in other AGN feedback simulations \citep[e.g.,][]{Yang2, Li}. While these parameters are given constant values in the runs described here, in Paper~II we link them to modeling of the black hole accretion disk.

The window function $\Psi$ determines the spatial extent of the jet. Defining cylindrical coordinates ($r, z$) with the $z$-axis oriented along the direction of the jet and having its origin at the AGN particle location, we use the following window function: 
\begin{equation}
\Psi (r,z) = \frac{1}{2\pi  \rjet^2}  \exp \left( -\frac{r^2}{2\rjet^2} \right)\frac{|z|}{\hjet^2}\ .
\end{equation}
The window function is set to zero outside the cylinder defined by $0 \leq |z| \leq \hjet$ and $r \leq \rjet$.
It is also normalized so that the total injected energy in the cylinder sums up to $\epsilon_{\rm f} \dot{M}_{\rm BH} c^2\Delta t$. In computing the value of $\Psi$ for a given zone we use the zone-center coordinates. 

For runs with jet feedback, in resetting the control surface region we want to exclude zones whose centers lie within the jet injection cylinder (both in computing the amount of material to remove and in applying the reset procedure). To achieve this, we move the jet inlet (the base of the cylinder) away from the $z=0$ plane to start at $z=\pm\rdep$. We have performed tests with different inlet offsets, but in those cases we found that often instabilities form below the injection cylinder and create outflows on the control surface away from the cylinder. Our intention is for the control surface region to serve as an inflow boundary except for where it intersects the injection cylinder, so we keep the jet offset at $\rdep$ in all these runs. 

\subsection{Tests of Jet Feedback}
\label{sec:jet_tests}

To test the jet injection model, we performed three-dimensional hydrodynamical simulations with the sink particle method and the jet model within a box of size 256~pc in the $xy$ plane and 512~pc along the $z$-axis. The grid was initialized to have uniform density and pressure values resembling ones at the center of the Perseus cluster \citep{Fabian06}. This is a crude approximation of a real cluster, but since the gravitational potential of the SMBH dominates at this scale, and the gas scale radius is much larger than the box, a uniform setup is justified. The electron density and pressure are taken to be $n_e = 0.05\ \rm cm^{-3}$ and $P = 0.67\ \rm keV\ cm^{-3}$, and the gas is initially at rest.  The ICM is treated as a fully ionized gas with an adiabatic index $\gamma = 5/3$. We use a point-mass gravitational potential corresponding to an SMBH mass of $10^9 M_{\odot}$ (we also do tests with various SMBH masses). With these values, the accretion radius is $\rb = 2.28{\rm\ pc}$ except for runs with varying SMBH masses. The jet axis is taken to lie in the $z$-direction.

Cooling is ignored for these tests, as the cooling time is larger than the time followed in the simulations. Additional consideration of microphysics processes in the ICM is left for later work. The JA runs use a uniform mesh, but in the other series, we use three levels of mesh refinement, with the finest level having a zone spacing $\Delta x$. For different resolutions, we keep the block structure the same, but vary the number of zones per block. The base grid contains 8 $\times$ 8 $\times$ 16 blocks. Run parameters for these tests are shown in Table~\ref{table_jet}. 

\begin{table*}
\begin{center}
  \caption{Table of test runs including jet feedback. In each case the ambient medium was set up to resemble the center of the Perseus cluster as described in the text.}
  \begin{tabular}{ll  c c   c  c c r }
    \hline
     Objective &
     Run Number & $M_{BH}/M_{\odot}$ & $\Delta x\ ({\rm pc})$ & $\rcs/\Delta x$ & $\rdep/ \Delta x$ & $\rjet/\Delta x$ & $\hjet/\Delta x$  \\ \hline
     Vary resolution of $\rb$ on uniform grid
     %& JA1 & 1.0 $\times 10^9$ & 0.25 & 8 & 4 & 2 & 2 \\ 
     & JA1 & 1.0 $\times 10^9$ & 0.5 & 8 & 4 & 2 & 2 \\ 
     & JA2 & 1.0 $\times 10^9$ & 1 & 8 & 4 & 2 & 2 \\
     %$& JA4 & 1.0 $\times 10^9$ & 1.33 & 8 & 4 & 2 & 2 \\
     & JA3 & 1.0 $\times 10^9$ & 2 & 8 & 4 & 2 & 2 \\ 
     & JA4 & 1.0 $\times 10^9$ & 4 & 8 & 4 & 2 & 2 \\  \hline
     Vary resolution of $\rjet$
     & JB1 & 1.0 $\times 10^9$ & 0.25 & 8 & 4 & 1 & 4\\ 
     & JB2 & 1.0 $\times 10^9$ & 0.25 & 8 & 4 & 2 & 4 \\ 
     & JB3 & 1.0 $\times 10^9$ & 0.25 & 8 & 4 & 3 & 4 \\ 
     & JB4 & 1.0 $\times 10^9$ & 0.25 & 8 & 4 & 4 & 4 \\  
     & JB5 & 1.0 $\times 10^9$ & 0.25 & 8 & 4 & 8 & 4 \\  \hline
     Vary resolution of $\hjet$
     & JC1 & 1.0 $\times 10^9$ &  0.25 & 8 & 4 & 2 & 2\\ 
     & JC2 (Same as JB2) & 1.0 $\times 10^9$ & 0.25 & 8 & 4 & 2 & 4 \\ 
     & JC3 & 1.0 $\times 10^9$ & 0.25 & 8 & 4 & 2 & 8 \\  
     & JC4 & 1.0 $\times 10^9$ & 0.25 & 8 & 4 & 2 & 16 \\  \hline  
     Vary $\rb$ relative to all
     & JD1 & 3.0 $\times 10^9$ & 0.25 & 8 & 4 & 2 & 2\\ 
     & JD2 (Same as JC1) & 1.0 $\times 10^9$ & 0.25 & 8 & 4 & 2 & 2 \\ 
     & JD3 & 3.0 $\times 10^8$ & 0.25 & 8 & 4 & 2 & 2\\  
     & JD4 & 1.0 $\times 10^8$ & 0.25 & 8 & 4 & 2 & 2 \\  \hline
     Test with $\alpha$-Bondi accretion rate
     & J0 & 1.0 $\times 10^9$ & 0.25 & 8 & 4 & 2 & 2\\ \hline
     \hline
  \end{tabular}
  \label{table_jet}
\end{center}
\end{table*}

\subsubsection{Resolution Dependence}

Here we consider the effect of spatial resolution on our feedback model (shown via runs in the JA series). The notion of convergence is complicated for subgrid models: as we consider smaller resolution elements, the ``physically appropriate'' subgrid model may change, and in the limit of infinite resolution we should not need a subgrid model at all. Self-convergence of a code never formally implies convergence to the correct solution, but when a subgrid model is involved we must also consider whether the model is valid at the scales under consideration. Keeping these considerations in mind, here we consider self-convergence over the range of zone spacings we might typically encounter in galaxy or galaxy cluster simulations. These spacings are much larger than the accretion disk size, so we do not expect the expanded suite of physics needed for global accretion disk simulations to apply.

Using our BHL tests as guidance, we have thus fixed the control surface and reset radii with respect to different resolutions, so as to have a better comparison for the relationship between the Bondi radius and the control surface radius. Runs JA1, JA2, and JA3 have $\rb > \rcs$, while the opposite is true for run JA4. 

\begin{figure}
    \begin{center}
    \includegraphics*[width=1.1\columnwidth]{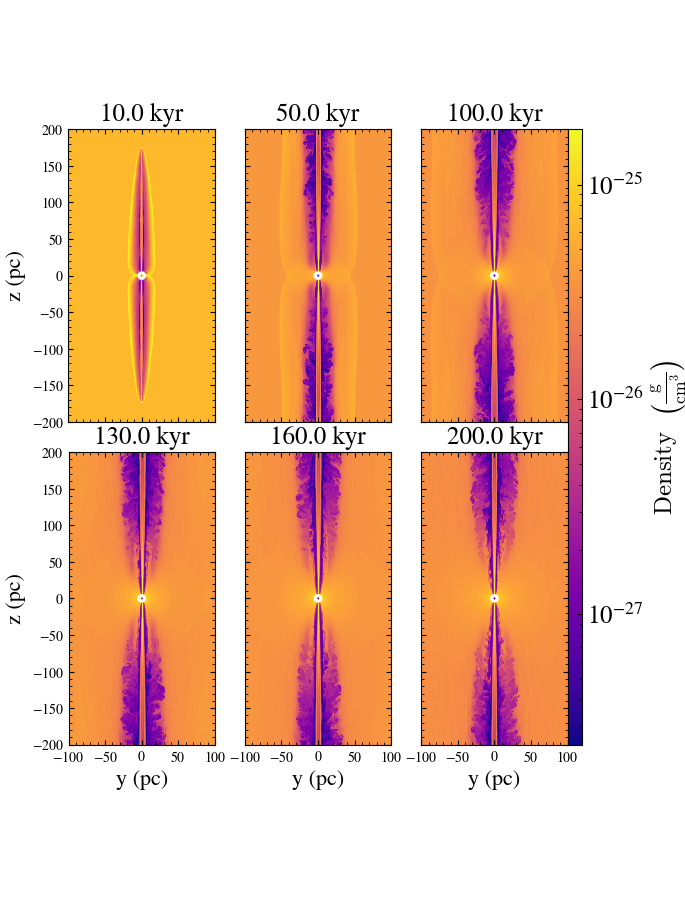}
    \caption{Density slices at different times for run JA1 with size of control surface $\rcs = 4$~pc, fixing the jet size at $\hjet = \rjet = 2 \Delta x$, where $\Delta x = 0.5\ {\rm pc}$. The white circle represents the control surface radius $\rcs$.}
    \label{slice0}
    \end{center}
\end{figure}

The time evolution of the most well-resolved run (with a uniform grid having resolution of $\Delta x = 0.5\ {\rm pc}$) is shown in Figure~\ref{slice0}. In this figure we show the $yz$ slice of the gas density field at $x = 0$ for different times in the simulation. The jets propagate along the $z$-axis, forming low density channels while inflating a cocoon of entrained gas. The jet reaches the edge of the box quickly (in less than 50~kyr), and the simulation converges to a steady state where the jet resembles a nozzle with secondary shock structures along the propagation axis. The secondary shock structures become more complex over time.

\begin{figure*}
    \centering
    \includegraphics*[width=2.\columnwidth]{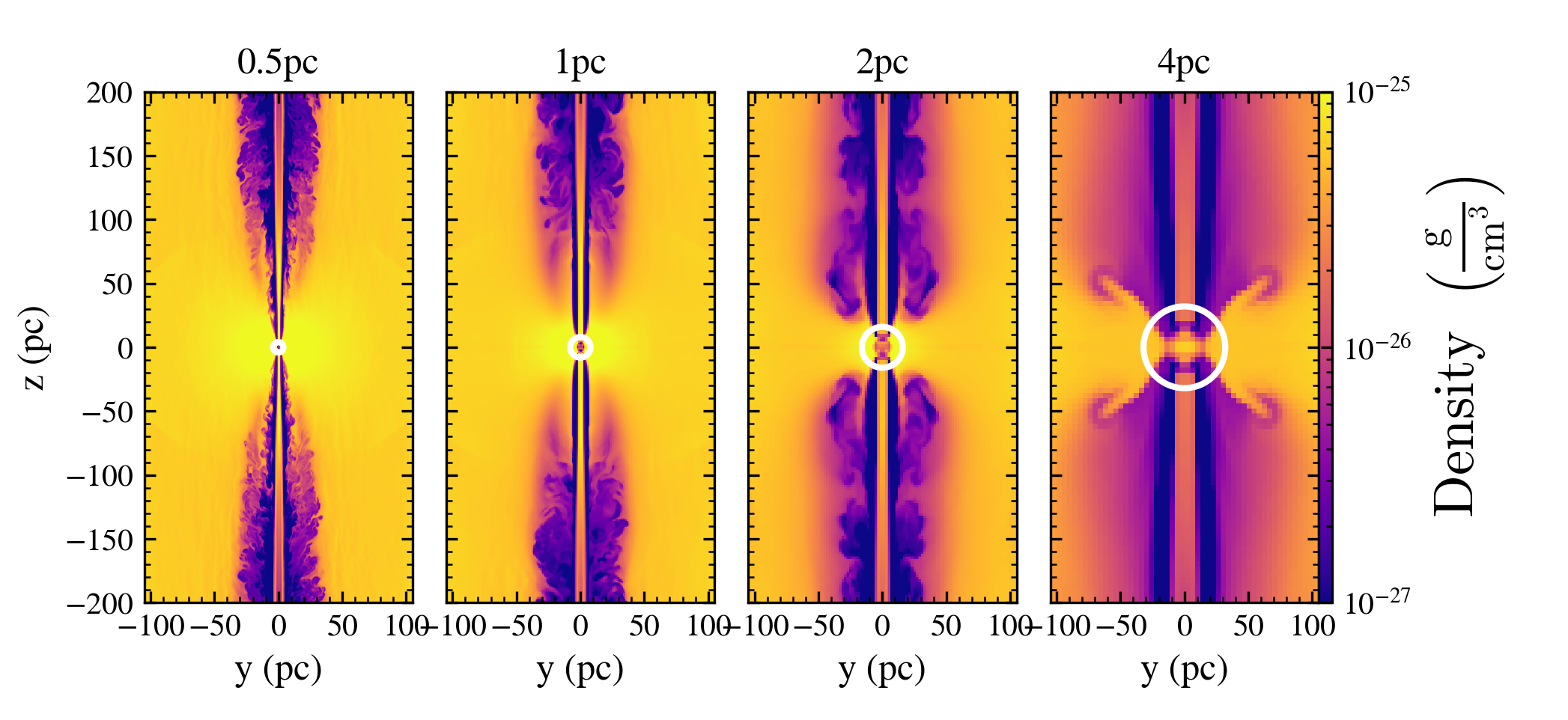}
    \caption{Gas density slice at $t = 0.2\ {\rm Myr}$ for JA runs.}
    \label{slice1}
\end{figure*}

The jet flow at a later time in the simulations with different resolutions is shown in Figure~\ref{slice1}. In the figure we see similar flow structures, and in all cases the jets reach the boundary of the domain. However, the structure of the gas flow along the jet axis varies significantly with resolution. For better-resolved runs, the shocked region is smaller and low density is observed along the jet axis. The secondary shock structures remain in horizontal distances close to the jet, but occur further away from the accretor vertically, and they do not seem to affect the flow through the control surface. A shock diamond is also observed. For poorly resolved runs however, the jet forms a high density channel and a backflow closer to the accretor.

With varying resolution in the JA series, the quantity that varies is $\rb/\Delta x$. Based on the jet structure we observe, we need $\rb > 2 \rcs$ in order to have a converged jet flow structure (corresponding to a resolution of 0.5~pc for this black hole mass). Since our tests run on a uniform grid, we are constrained by computing power, but for later runs, we use an adaptive grid to reach a higher resolution of 0.25 pc. 

\begin{figure*}
    \begin{center}
    \includegraphics*[width=1.\columnwidth]{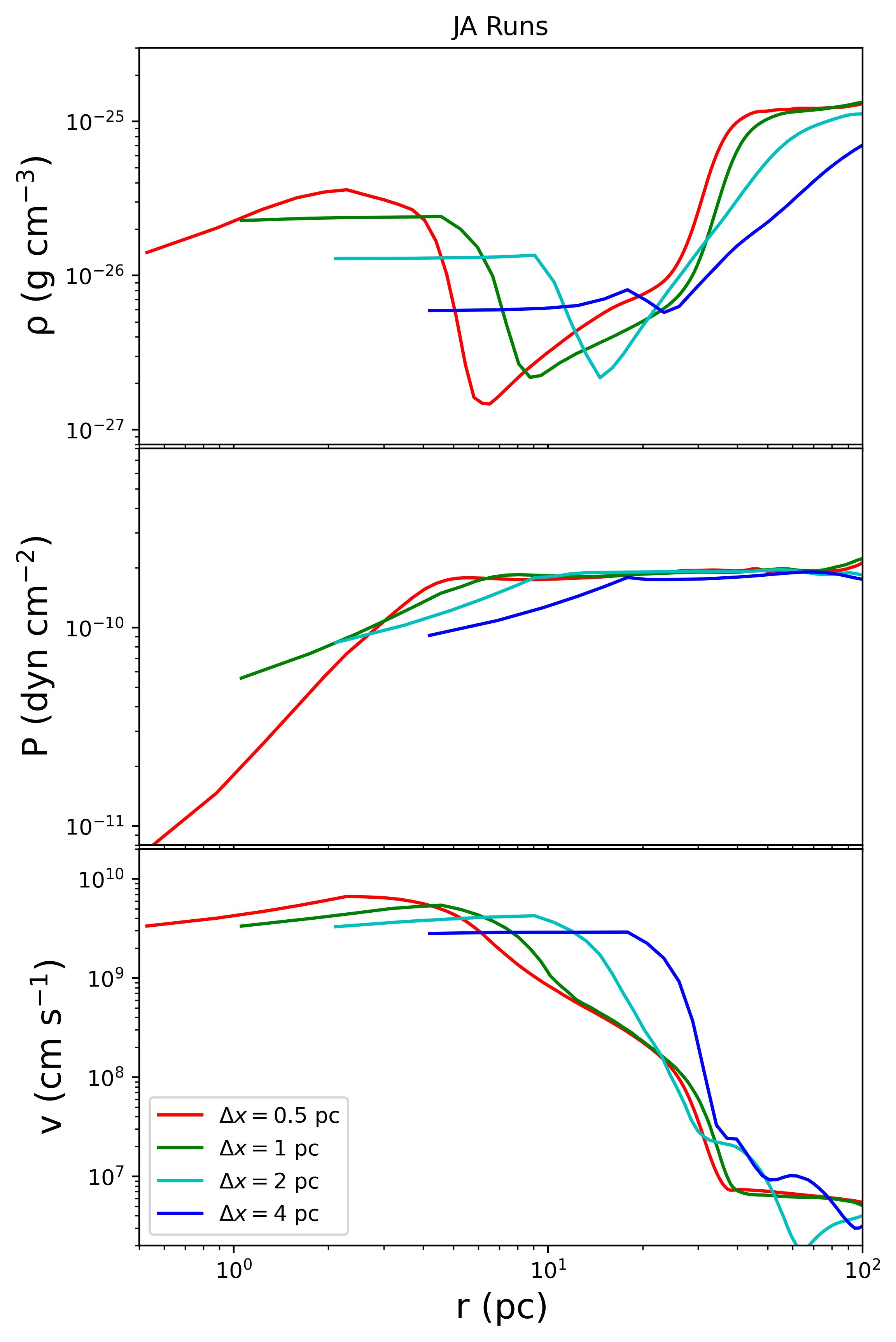}
    \includegraphics*[width=1.\columnwidth]{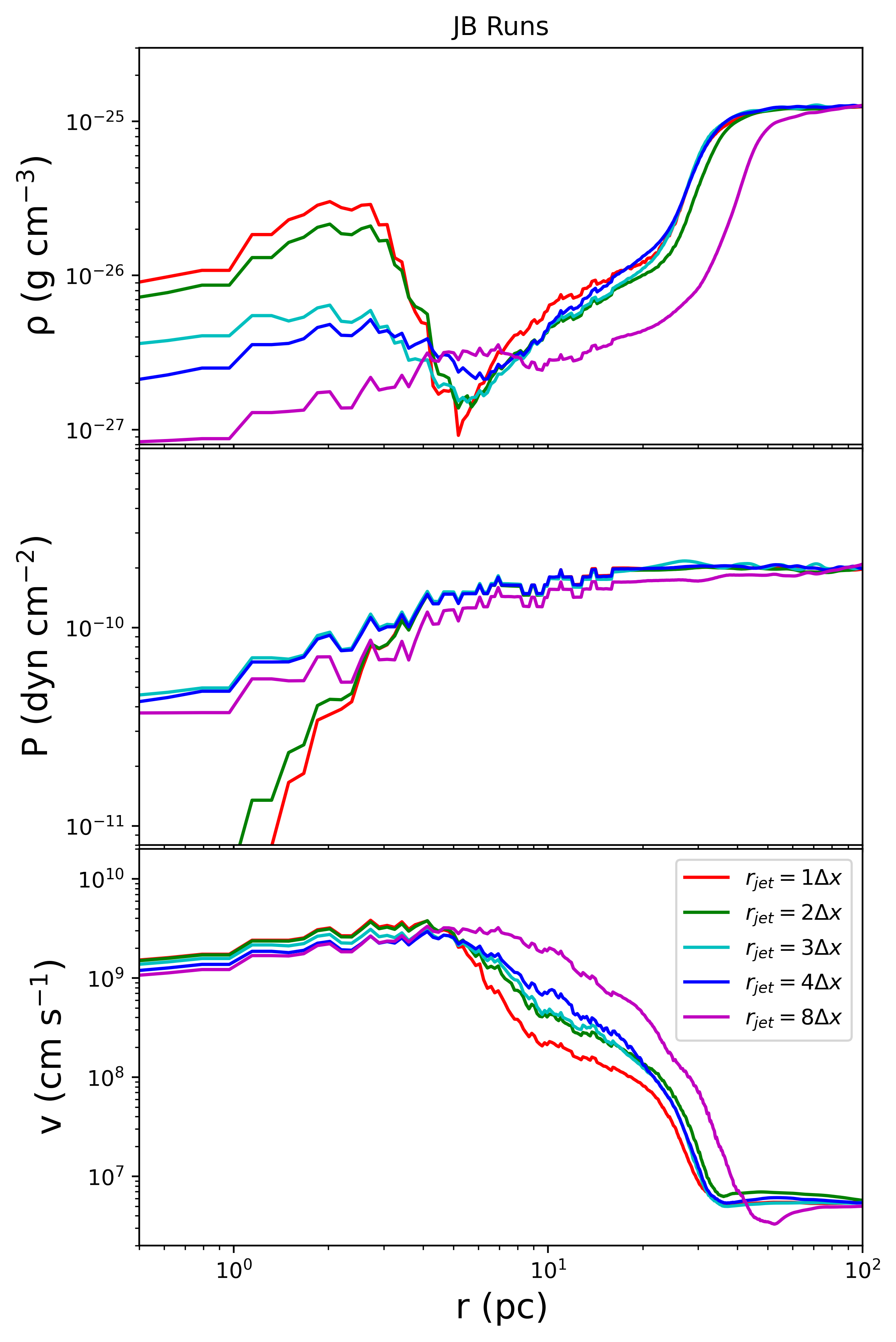}
    \caption{Azimuthally and longitudinally averaged density, pressure, and velocity profiles at $t = 0.2\ {\rm Myr}$ for different jet runs.  Left: with varying resolution, where control surface and reset radius are scaled with the resolution (JA runs). Right: fixing resolution and varying jet radius (JB runs). }
    \label{profile1}
    \end{center}   
\end{figure*}

To examine the jet structure in more detail, in Figure~\ref{profile1} we plot azimuthally-averaged gas profiles in slices perpendicular to the jet axis at a late time for the runs. Since the jet lies along the $z$-axis, for each plot we choose a set of 100 $z$ values, compute the azimuthally averaged density, pressure, and velocity magnitude versus distance $r=\sqrt{x^2+y^2}$ between 100~pc and $L_{\rm box}/2$ from the jet axis for each $z$ value, and then average the profiles over the $z$ values. To avoid boundary effects closer to the reset radius and the domain boundary, we take only the z values larger than the reset radius and up to 250 pc. The left panels in Figure~\ref{profile1} show results for the JA runs, varying key radii together with resolution. 

In the density profiles of the JA runs, we see a drop from the central value for all cases. The poorer the resolution, the larger the turn-over radius, which is consistent with Figure~\ref{slice1}, where poor resolution runs have jets that cover a larger region. However, more poorly resolved runs also show a lower average density outside the jet. Most of the jet material is in pressure equilibrium with the ambient medium, with a drop towards the central area closer to the spine of the jet. The differences for larger radii among runs are negligible. The jet flow speed at its maximum also remains consistent, reaching over \newtext{ $\sim 10^4 {\rm\ km\ s}^{-1}$} and dropping more than two orders of magnitude for the gas farther away. However, well-resolved runs exhibit a more rapid drop of velocity, since cells around the jet edge are under-resolved.

\begin{figure*}
    \begin{center}
    \includegraphics*[width=1.\columnwidth]{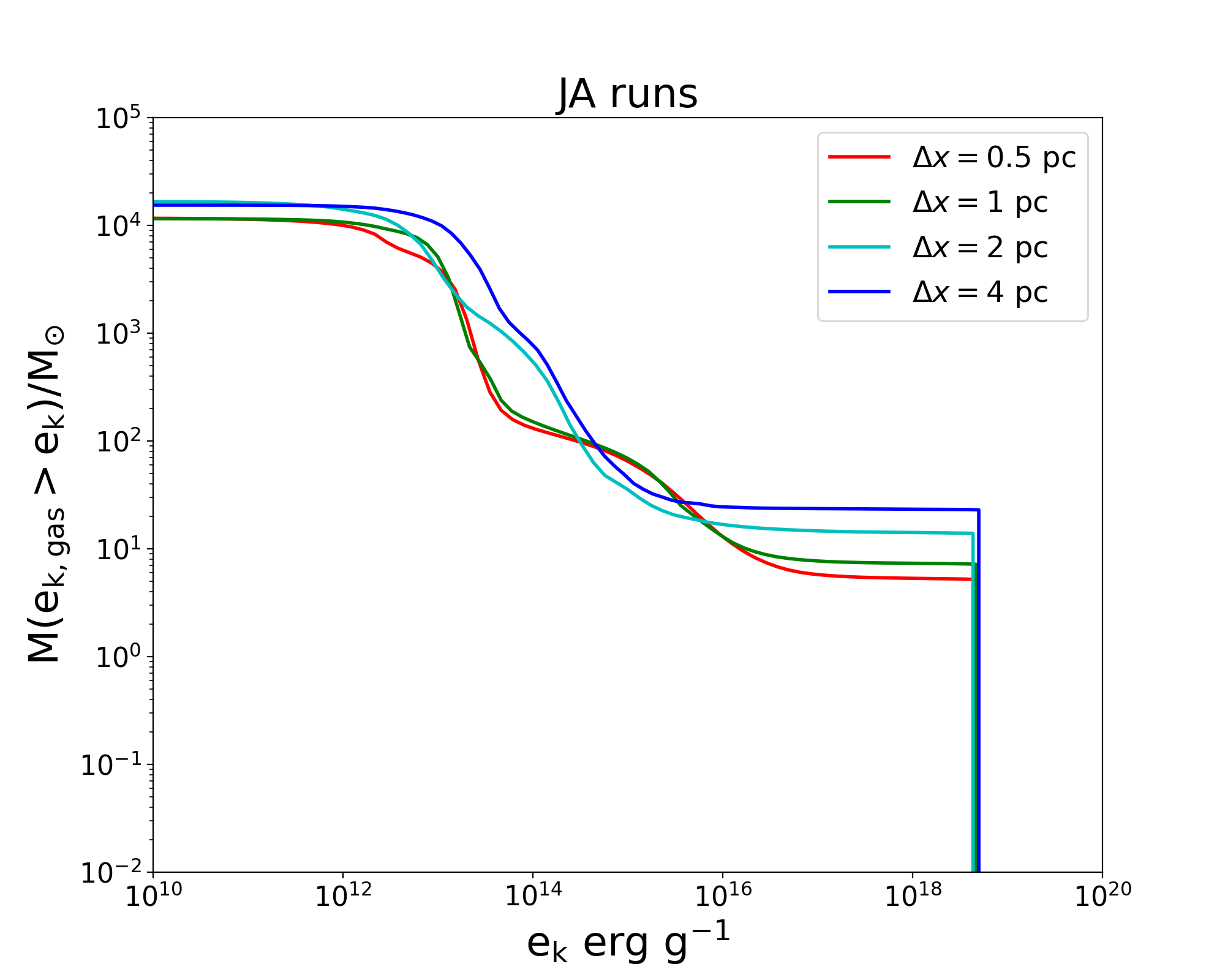}
    \includegraphics*[width=1.\columnwidth]{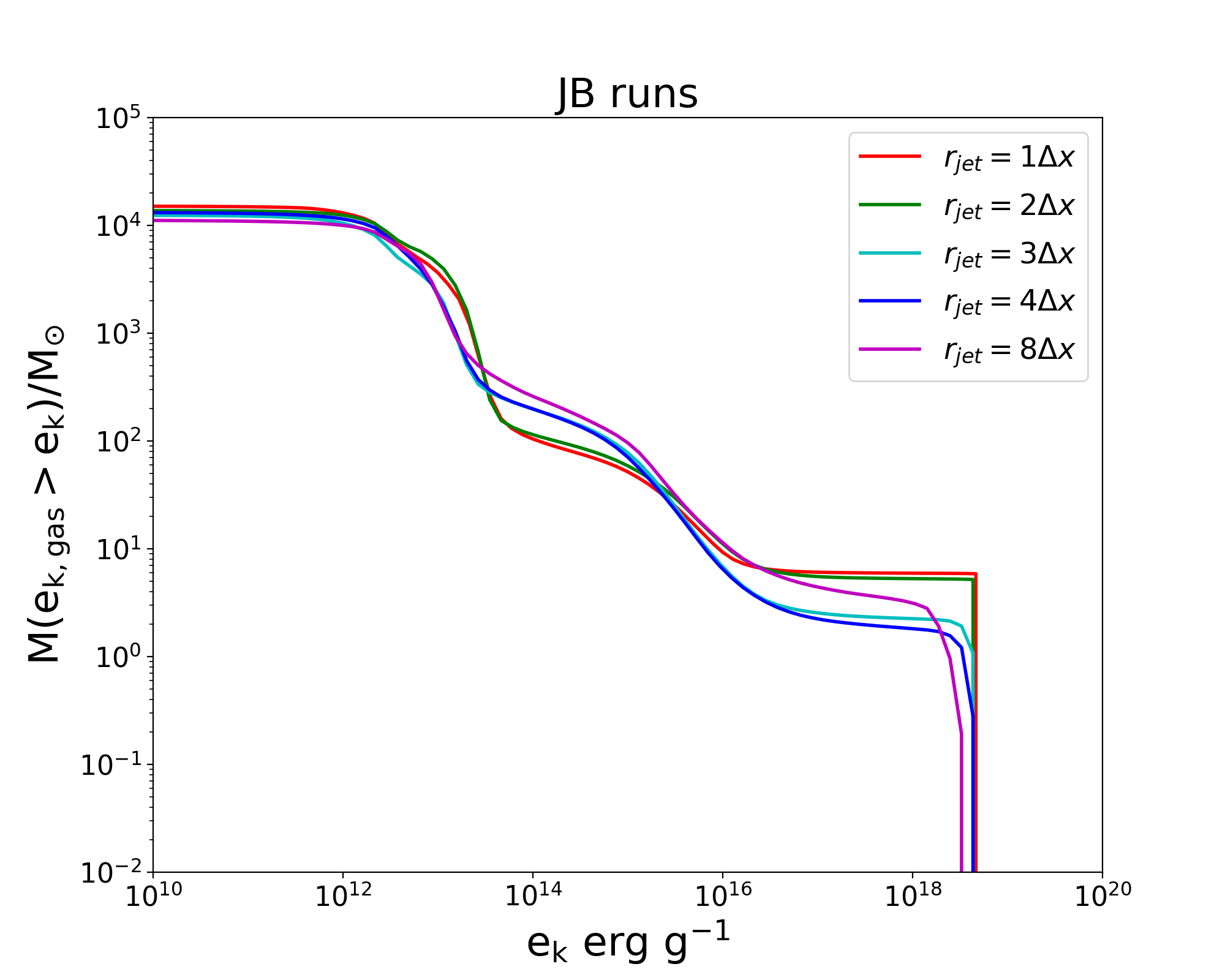}\\
    \includegraphics*[width=1.\columnwidth]{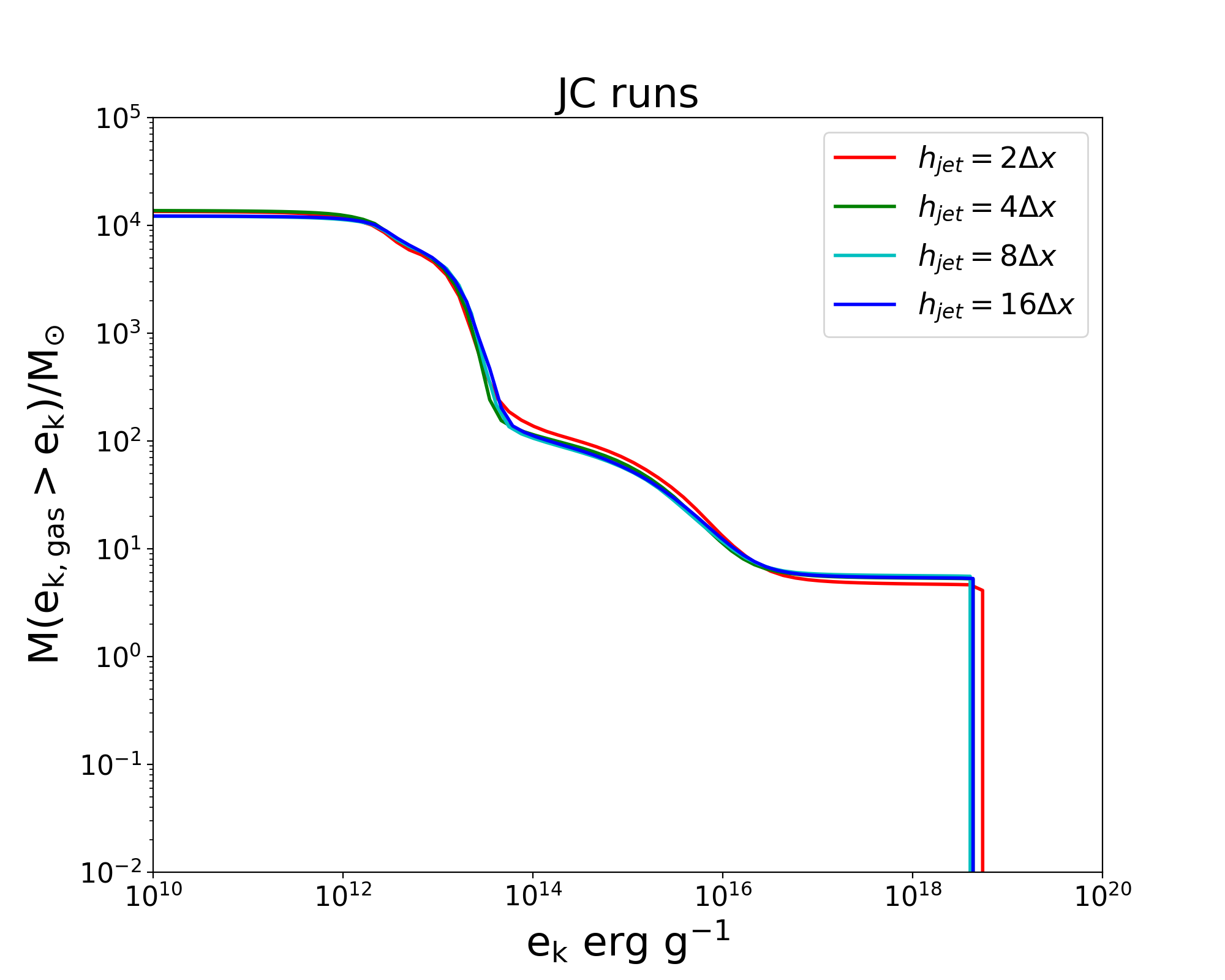}
    \includegraphics*[width=1.\columnwidth]{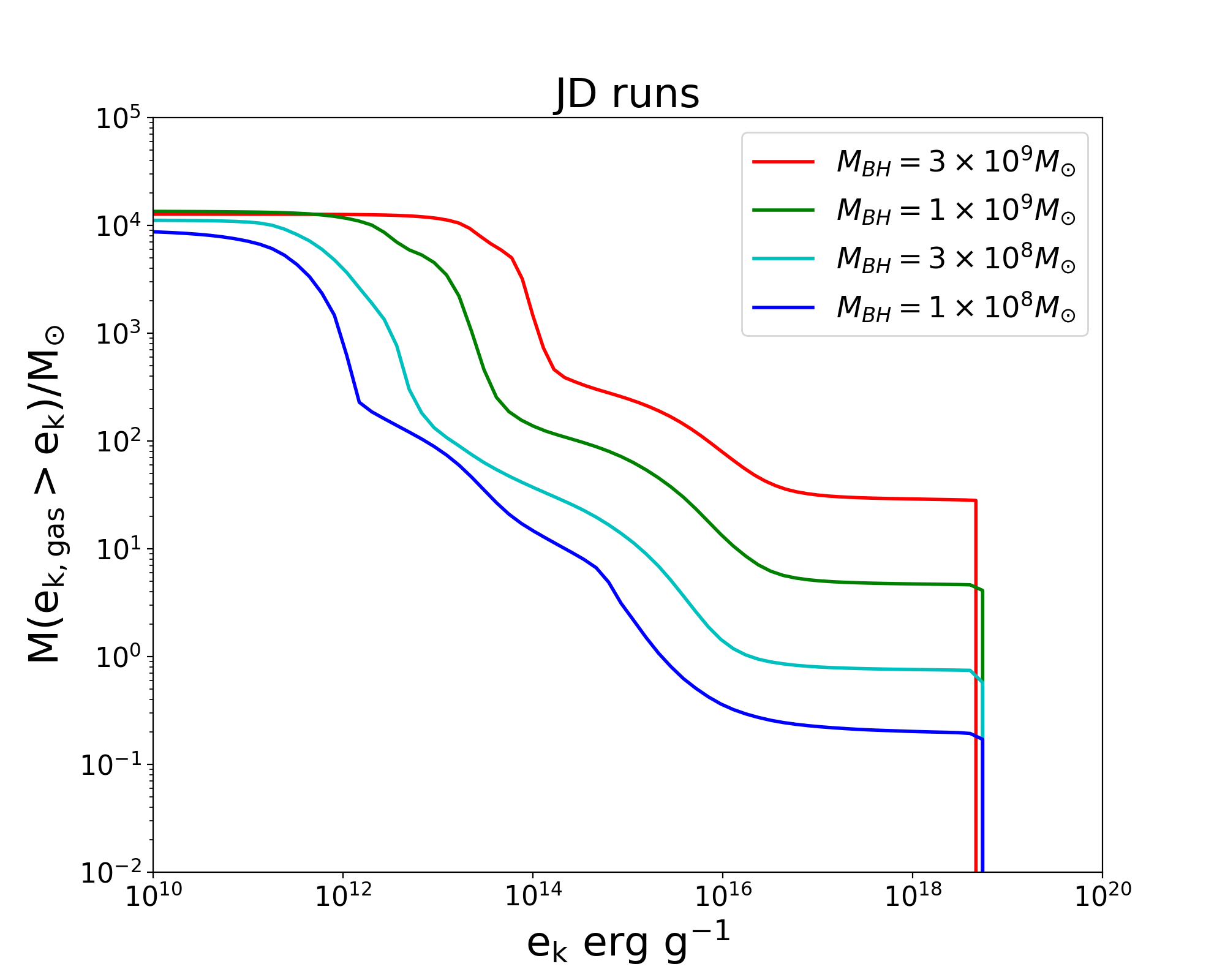}
    \caption{Cumulative kinetic energy distribution of entrained material for different runs (JA, JB, JC and JD).
    }
    \label{ment}
    \end{center}
\end{figure*}

Using the cumulative distribution $M(> e_k)$ of specific kinetic energies $e_k$, we can determine the mass of gas entrained by the jet. 
In Figure~\ref{ment} we show this distribution for the different jet runs. In all these curves there exist three main components. At low $e_k$ we have undisturbed material at or near rest. The step on the left (typically at or below $10^{14}{\rm\ erg\ g}^{-1}$) separates this gas from the entrained material, while the step on the right (around $10^{16}{\rm\ erg\ g}^{-1}$) marks the jet itself. The sharp drop to zero (above $10^{18}{\rm\ erg\ g}^{-1}$) marks the spine of the jet. The mass of entrained material is then the difference between the value of $M(>e_k)$ at the base of the left step and its value at the base of the right step.

The JA runs show convergence for $M(>e_k)$ for $\Delta x \le 1{\rm\ pc}$, with $\sim 200-300M_\odot$ of entrained material, suggesting that $\rb > \rcs$ is needed for convergence of this statistic. The JB and JC runs show some sensitivity to the jet injection radius $\rjet$, with convergence for $\rjet \le 2\Delta x$, but almost no dependence on the jet injection height $\hjet$. The JD runs, which vary the Bondi radius relative to the other quantities, show a steady decrease in entrained mass with increasing $\rb$, though the mass that is entrained moves faster. The lowest-mass run (JD4) does not resolve $\rb$, and as a result $M(> e_k)$ shows some qualitative differences with the runs that do resolve $\rb$.

\begin{figure*}
    \begin{center}
    \includegraphics*[width=1.\columnwidth]{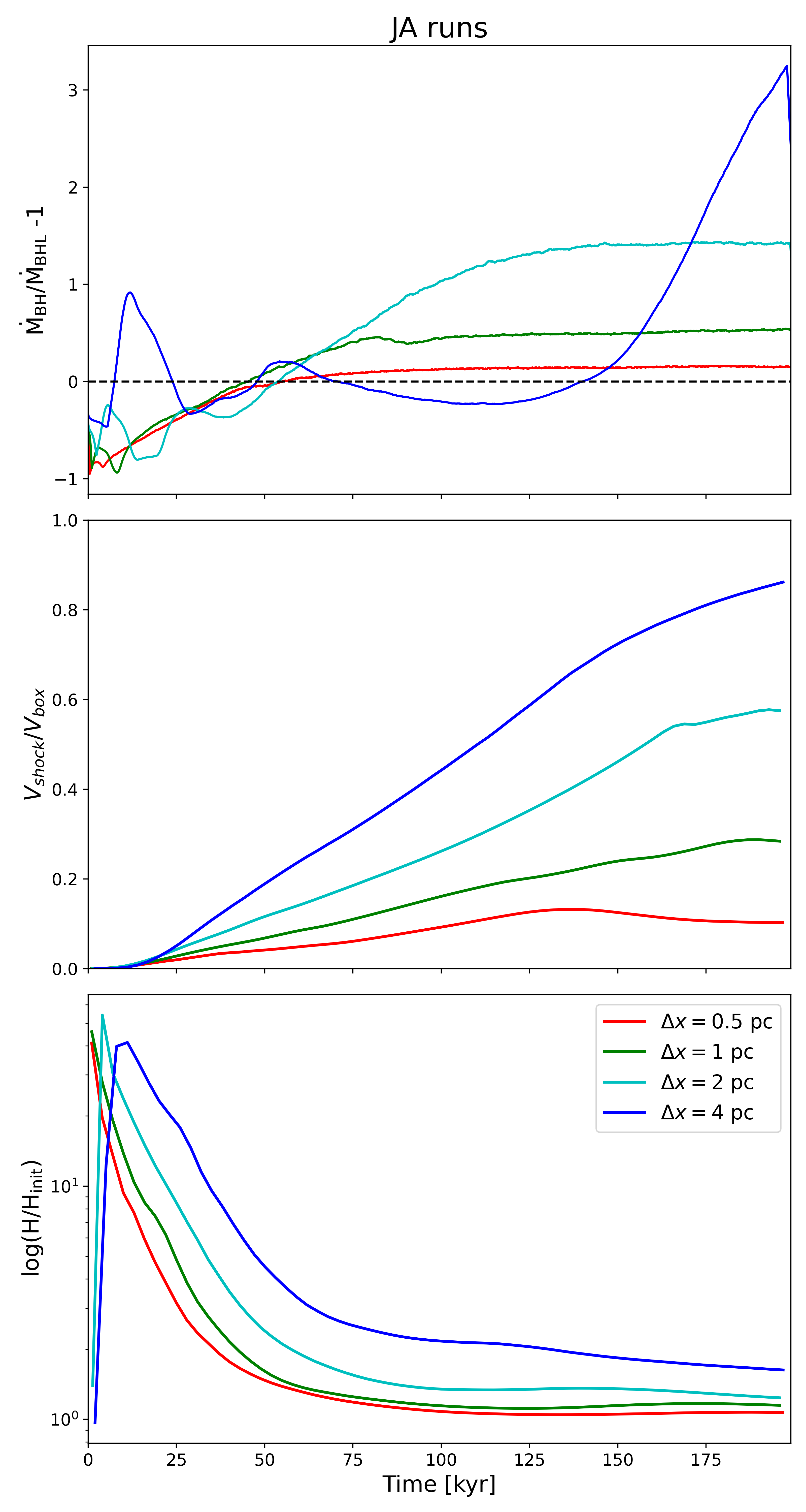}
    \includegraphics*[width=1.\columnwidth]{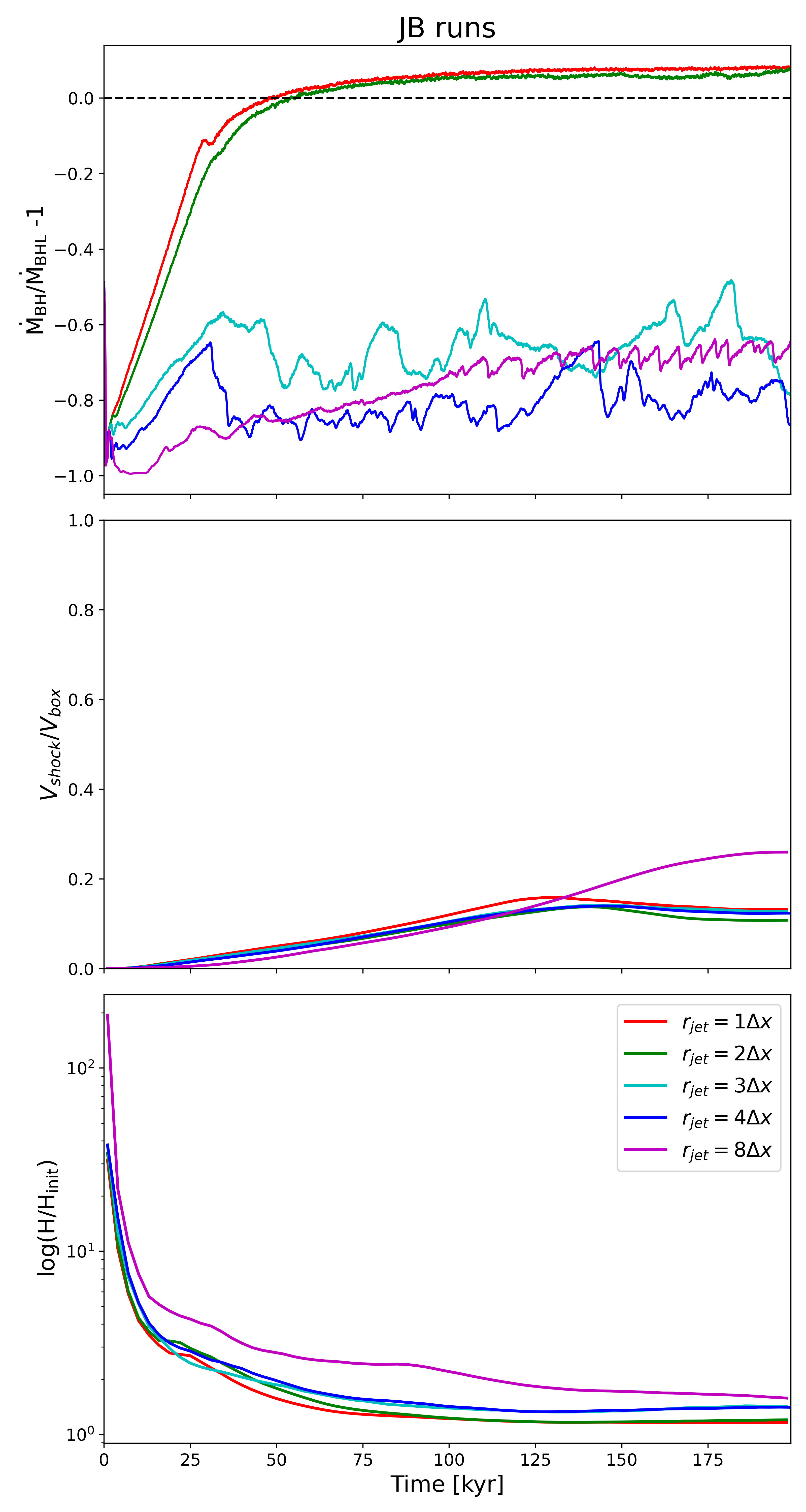}
    \caption{Evolution of accretion rate, shocked volume and enthalpy for the JA and JB runs.}
    \label{macc1}
    \end{center}
\end{figure*}

\begin{figure*}
    \begin{center}    
    \includegraphics*[width=1.\columnwidth]{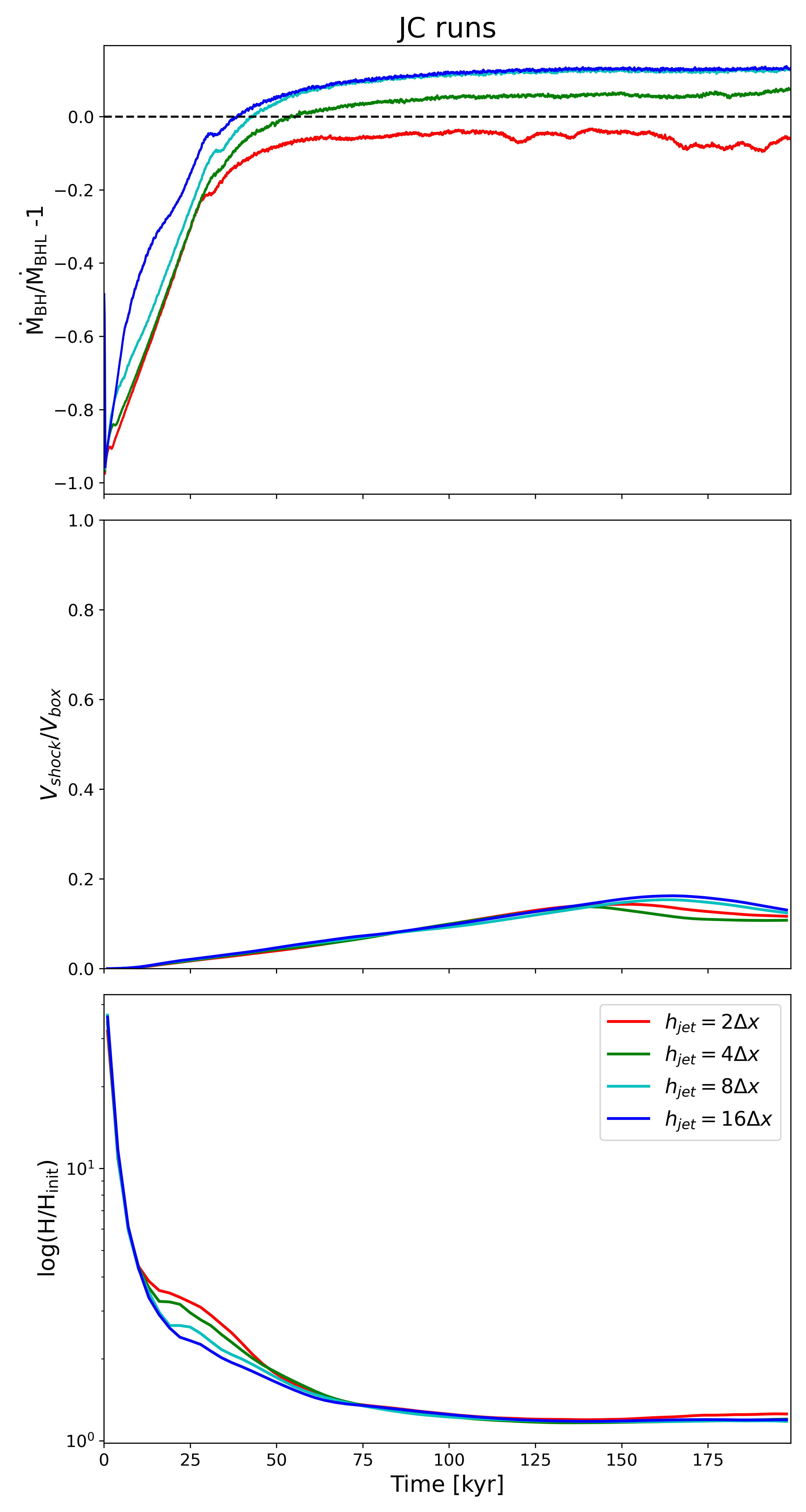}
    \includegraphics*[width=1.\columnwidth]{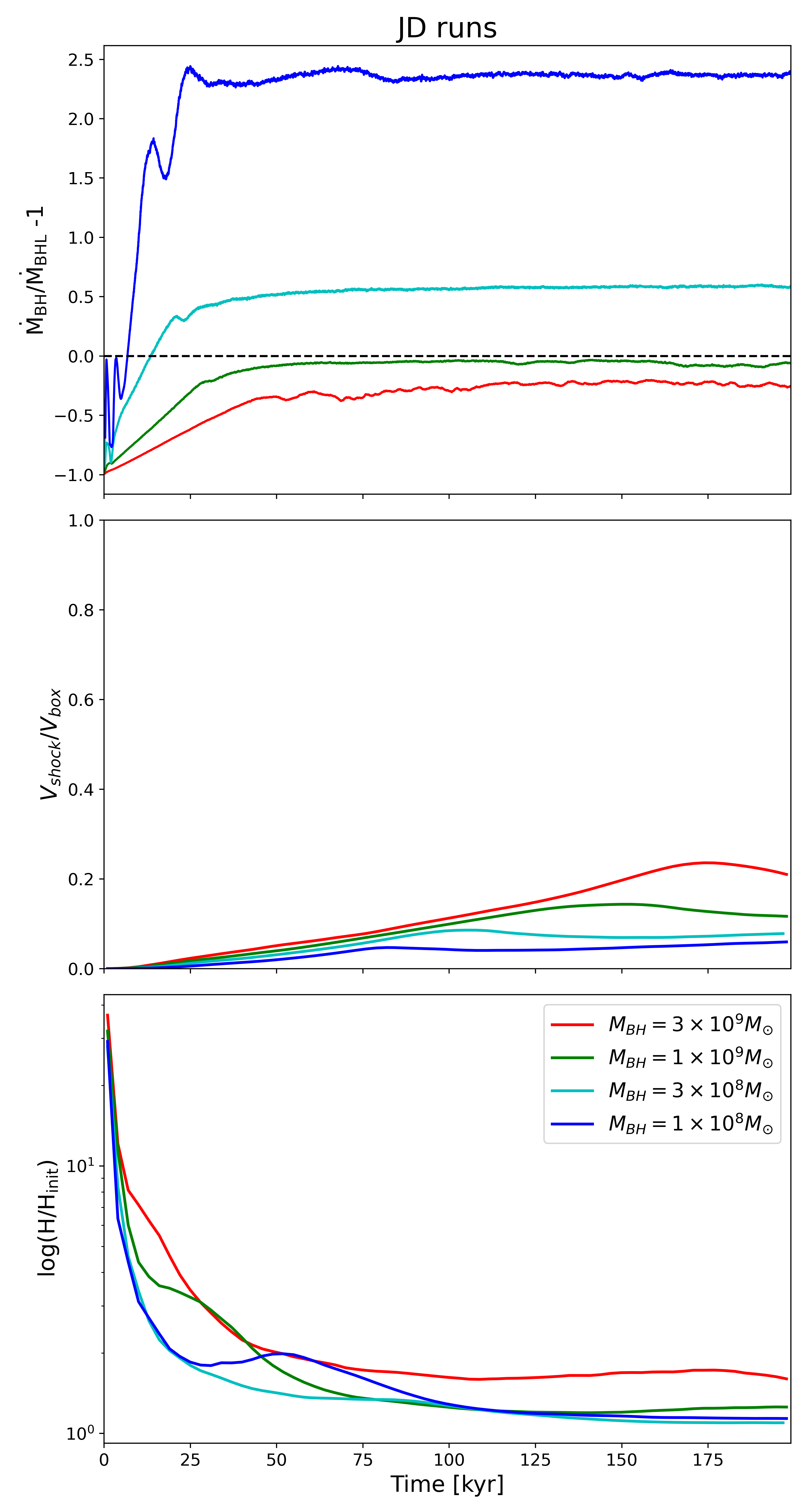}
    \caption{Evolution of accretion rate, shocked volume and enthalpy for the JC and JD runs.}
    \label{macc2}
    \end{center}   
\end{figure*}

The time evolution of the accretion rate, shocked volume, and enthalpy are shown in Figures~\ref{macc1} and \ref{macc2}. The accretion rate curve here is smoothed using a linear convolution with a sequence of length 50, i.e.\ smoothing every 50 data points. For visual inspection, the accretion rate is scaled to the Bondi-Hoyle-Lyttleton accretion rate. The shocked volume is calculated by summing up the volume of cells with a higher than ambient entropy $P/\rho^{5/3}$ in the domain. To allow for numerical errors, we set this threshold with a 3\% error bar (which means if the entropy is higher than 97\% of the ambient value, it is considered within the shocked region). The enthalpy is scaled relative to the ambient. For the latter two quantities, we do not include data inside the accretion radius or the jet inlet.

Despite accretion starting out differently, from $t \sim$ 50 to 80 kyr onward, the accretion rate in most runs reaches a steady state where the accretion and feedback rate appear to stay relatively constant over time. For high resolution runs, the accretion rate stays around the Bondi-Hoyle-Lyttleton rate, but for lower resolutions, the accretion rate continues to increase, leading to an over-estimation. For the lowest resolution, however, there is a temporary under-estimation of accretion rate followed by accretion not able to reach steady state configuration. This is connected to the development of the jet feedback. Since fluctuations of the gas flow between different cells are better captured with increasing resolution, there is a more obvious self-regulation effect -- more gas inflow through the control surface leads to stronger jet feedback, which pushes away more gas around the SMBH, and thus the accretion rate drops, leading to a fluctuation. For poorly resolved runs, despite the presence of the jet, it is also not able to regulate the inflow of the gas back onto the SMBH. In general, with resolution better than or equal to 1~pc, we are able to see the feedback transitioning into a more steady state configuration.

The volume of shocked material also increases as the jet cocoon expands, as seen in the middle panel of the figure. Eventually, the shocked volume fills up a large fraction of the grid. This value varies with resolution, as for the most well-resolved run, the shock only fills up about 10\% of the grid, while for the run with resolution 2~pc, this number goes to about 60\%, and 80\% of the grid for poorest resolutions. This is partly due to the effect where, the simulation transitions into a more confined configuration for the jet for well-resolved runs. As turbulence increases within the jet region, there is more mixing of high-entropy and low-entropy material; thus some of the cells are not counted towards the shock because of the low entropy. After decreasing the entropy threshold (adding the 3 \% error bar), the shock volume increases for all times in the simulation. Overall, the time evolution behavior shows consistency between shock structure and jet power, and further the accretion rate onto the SMBH. The enthalpy also demonstrates a behavior that is consistent with the above conclusions. While enthalpy increases initially a the jet is filling up the simulation domain, higher resolution runs show jets filling up quicker and the enthalpy decreases afterwards. In the end, enthalpy is lower for higher resolution runs due to a more confined jet. 

Based on the above observations, we conclude that increasing resolution can lead to a better convergence of the jet feedback. The cell size of 1~pc (0.5$
\rb $) is the least acceptable resolution to achieve high fidelity in these jet feedback simulations. This means the control surface radius also needs to be less than half of $\rb$ for more physical runs. 

\subsubsection{Size of Jet Injection Region}
\label{sec:vary_size}

The height $\hjet$ and radius $\rjet$ of the jet injection cylinder are adjustable numerical parameters. Ideally the radius should reflect the typical radius of an AGN jet at the distance represented by the control surface radius. However, jets observed using very long baseline interferometry (VLBI) are still highly collimated at the Bondi radius and may have transverse radii of only a few percent of the Bondi radius at that distance \citep{BlandfordARAA}. Thus for the jet injection cylinder to be physically correct, we would require the control surface (which is typically about $\rb$ or $\rbhl$) to be impractically well-resolved. At best we can make the injection cylinder about two zones across. Also, the cylinder radius cannot be larger than the control surface radius $\rcs$, or else accretion will only be allowed from the sides.

Unlike the radius of the injection cylinder, its height does not have a physical correlation, since by means of the cylinder we are attempting to impose an inflow boundary condition (i.e. directed from the control surface onto the grid outside it) that determines the mass and energy flux on a two-dimensional surface. The cylinder height must be at least $\rcs$ in order to have an effect on gas outside the control surface, and to avoid nonphysically forcing the solution there it should not be much larger than $\rcs$.

\begin{figure*}
    \centering
    \includegraphics*[width=2.\columnwidth]{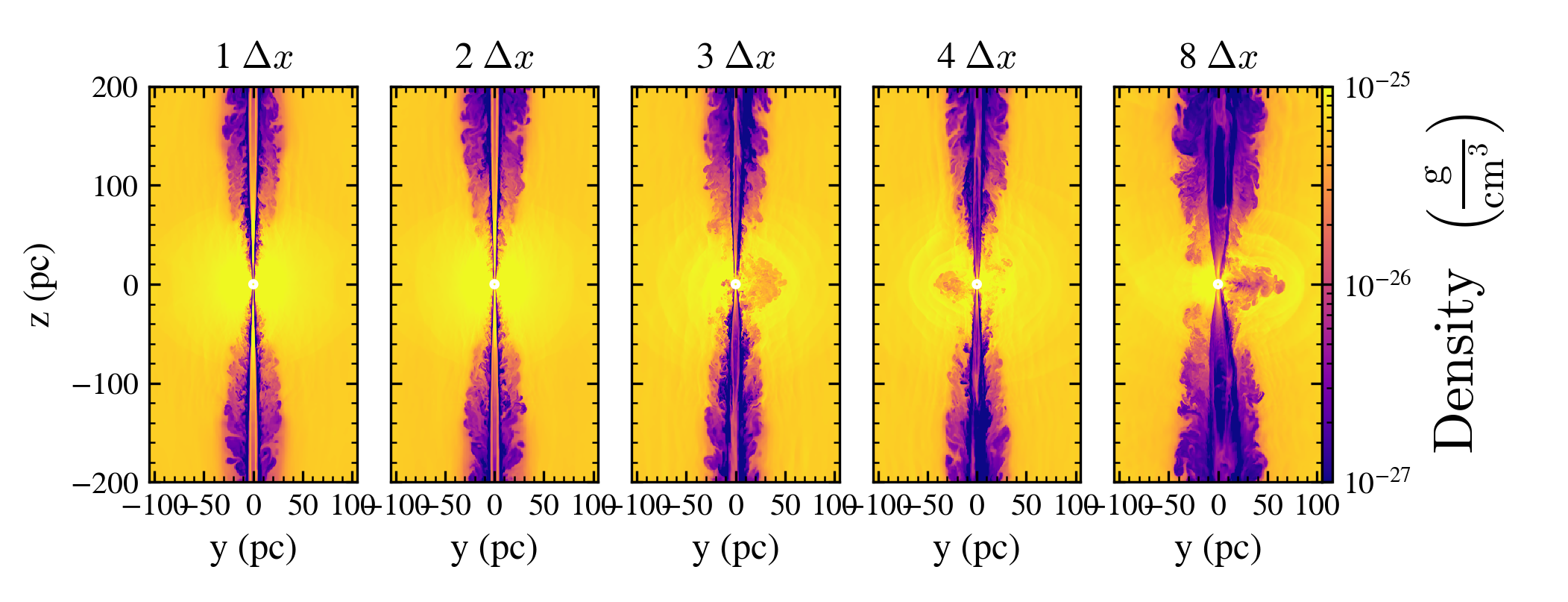}
    \includegraphics*[width=2.\columnwidth]{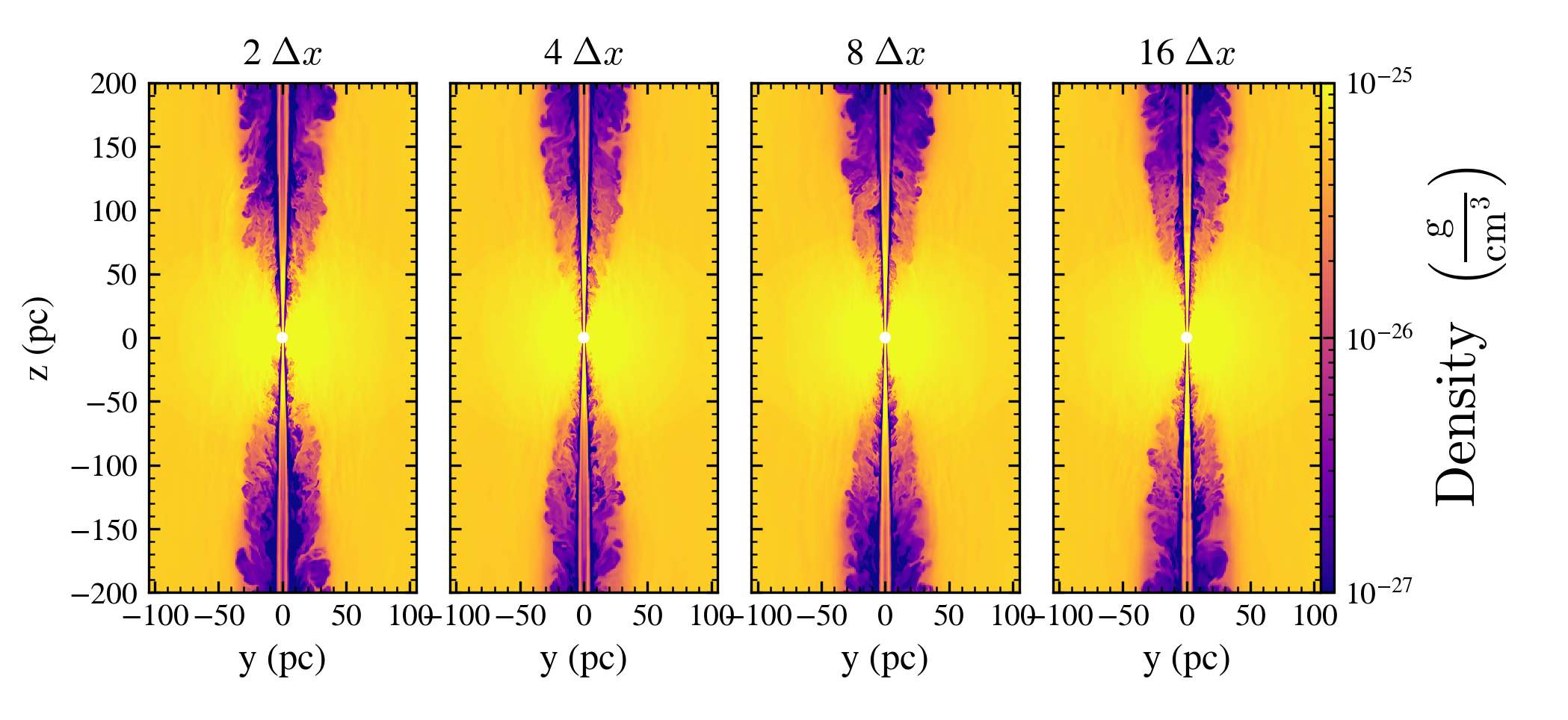}
    \caption{Gas density in the $yz$ plane at $t=0.2$~Myr for runs with fixed control surface size 
    and resolution but varying radius and height of the jet injection cylinder. Top panel shows runs varying $\rjet$ (JB runs) and bottom panel shows runs varying $\hjet$ (JC runs). }
    \label{slice2}
\end{figure*}

Given these constraints, we conducted runs for jet feedback with different values of $\hjet$ and $\rjet$ to explore the sensitivity of the feedback behavior to these parameters. The runs relevant for this section are the JB and JC runs. These runs used a well-resolved control surface with $\Delta x = 0.25\ {\rm pc}$. We kept the control surface radius $\rcs = 8 \Delta x = 2\ {\rm pc}$ and reset radius $\rdep = 4 \Delta x = 1\ {\rm pc}$. The JB runs vary jet radius, keeping $\hjet = 4 \Delta x$, while the JC runs keep $\rjet = 2 \Delta x$ and vary the height of the jet. Slice plots showing the typical jet structure in each run are shown in Figure~\ref{slice2}. 

\begin{figure*}
    \centering   
    \includegraphics*[width=1.\columnwidth]{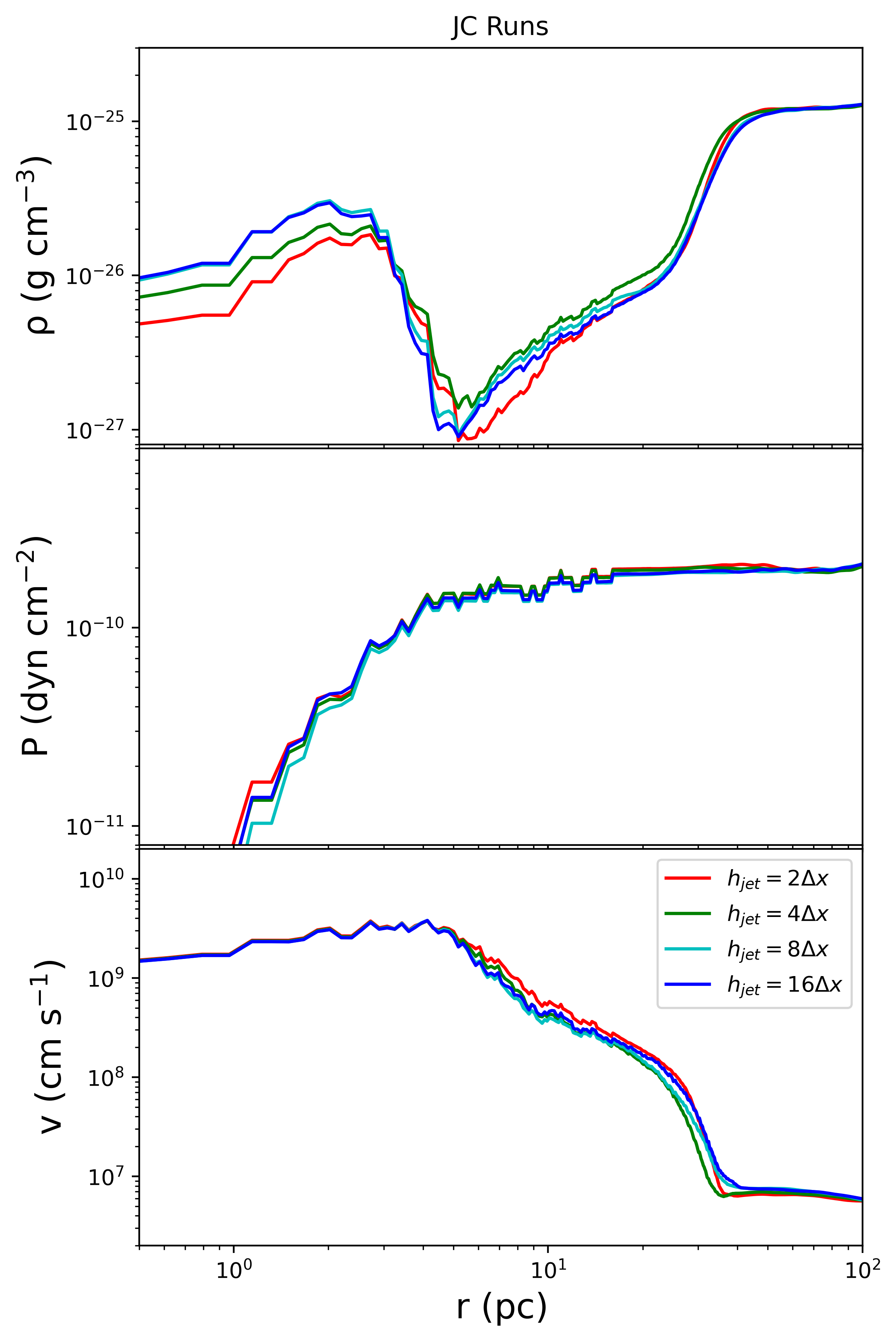}
    \includegraphics*[width=1.\columnwidth]{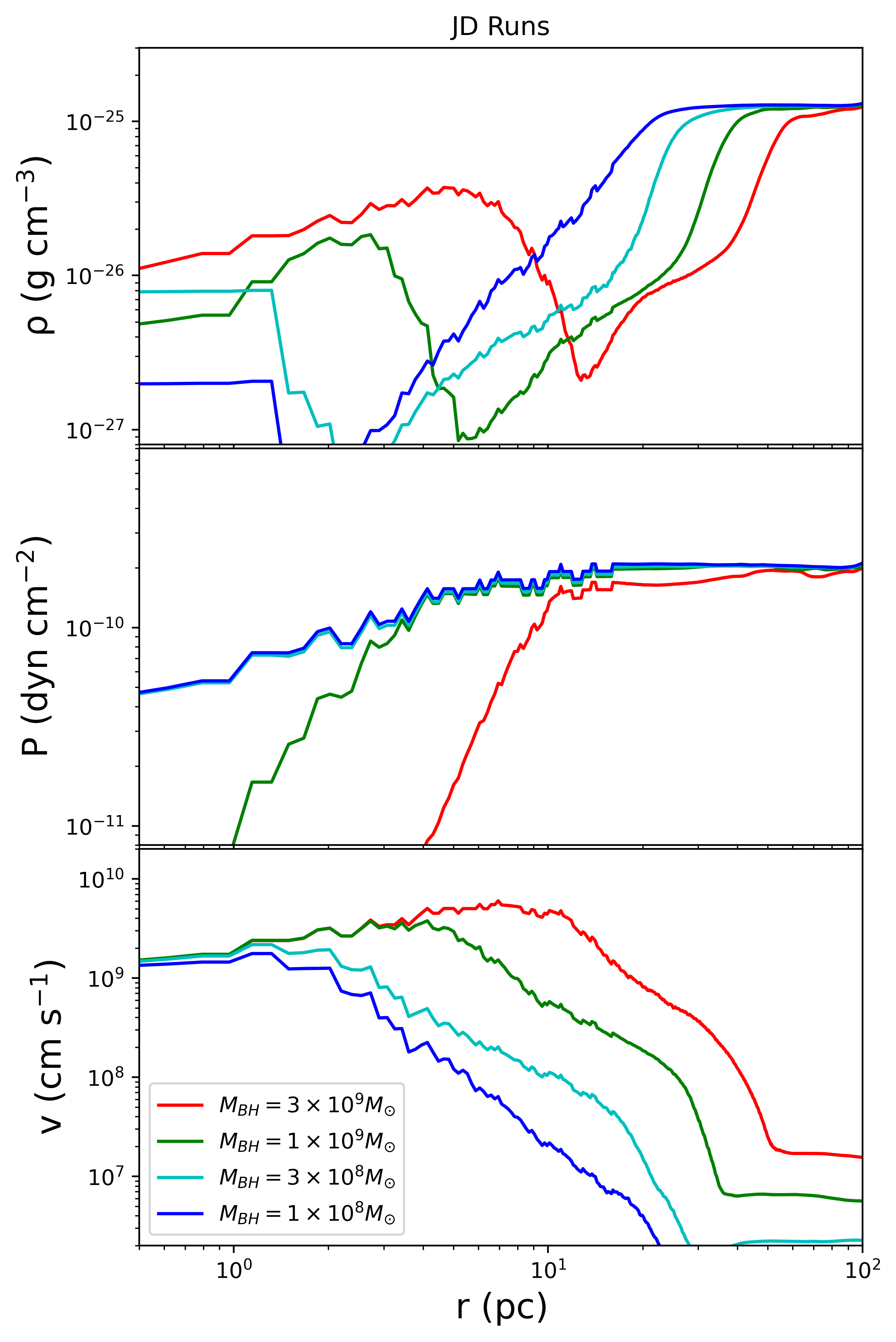}
    \caption{Figure \ref{profile1} continued. Left: With fixed resolution and jet radius, varying jet injection height (JC runs). Right: Varying the mass of the central BH and fixing other parameters (JD runs). }
    \label{profile2}
\end{figure*}

Our tests show that $\rjet$ is the more important of the two parameters. Regardless of $\hjet$, the narrower jets develop a dense spine with a nozzle structure and side shocks. The turbulence within the jet region can vary, but the eventual size of the shocked region varies little among the runs. From the azimuthally and longitudinally averaged profiles of density, pressure, and velocity (Figure~\ref{profile2}), we can also see that the pressure and jet velocity do not vary significantly with the height of the jet. The density in the jet spine varies by up to a factor of two with changing $\hjet$. The kinetic energy distribution (seen in Figure~\ref{ment}), is also virtually independent of $\hjet$. The accretion rate and shocked volume at late times (Figure~\ref{macc1} and \ref{macc2}) also show that varying the width of the jet injection cylinder has a stronger effect than varying its height. At late times, smaller jet heights can result in smaller shocked volume, consistent with what is observed in the azimuthally averaged profile. In general, we can conclude that the results are relatively insensitive to the height of the jet injection region.

The results are different with varying jet width, however. If $\rjet>\rdep/2$, the jet injection procedure \newtext{produces a combination of effects that lead to mass ejections perpendicular to the jet direction. First, a wide jet injection cylinder constricts infalling gas, producing a Venturi effect: material infalling from the side speeds up and develops lower pressure, drawing more material from the side. Second, since the velocity is not modified in the resetting region, the gas remaining after resetting exits the region at high speed, shocking against material falling in from the opposite direction. The result is a pseudo-jet whose direction is determined by random fluctuations in the flow field around the black hole.}. This unphysical behavior places an upper limit on acceptable values of $\rjet$\newtext{; to avoid it, the injection cylinder must not overhang the resetting sphere by more than a mesh zone. We also observe that} narrower jets form low-density central channels, while wider jets inflate shocks and cocoons with significant size. Their accretion rate and shocked volume show more variation as well, with the smaller injection cylinder allowing less accretion and producing a more slowly expanding transverse shock. Narrower jets also entrain significantly less material.

Based on the above results, a narrow injection cylinder achieves the most realistic results. Observationally, we also know that the jets are narrower even than $0.25\rcs$ (in this case 0.5 pc). To achieve a realistic accretion rate according to the simulations, we require $\rjet$ to be at least two zones across, and no larger than $\rdep/2$.

\subsubsection{SMBH Mass}
\label{sec:vary_mass}

\begin{figure*}
    \centering
    \includegraphics*[width=2.\columnwidth]{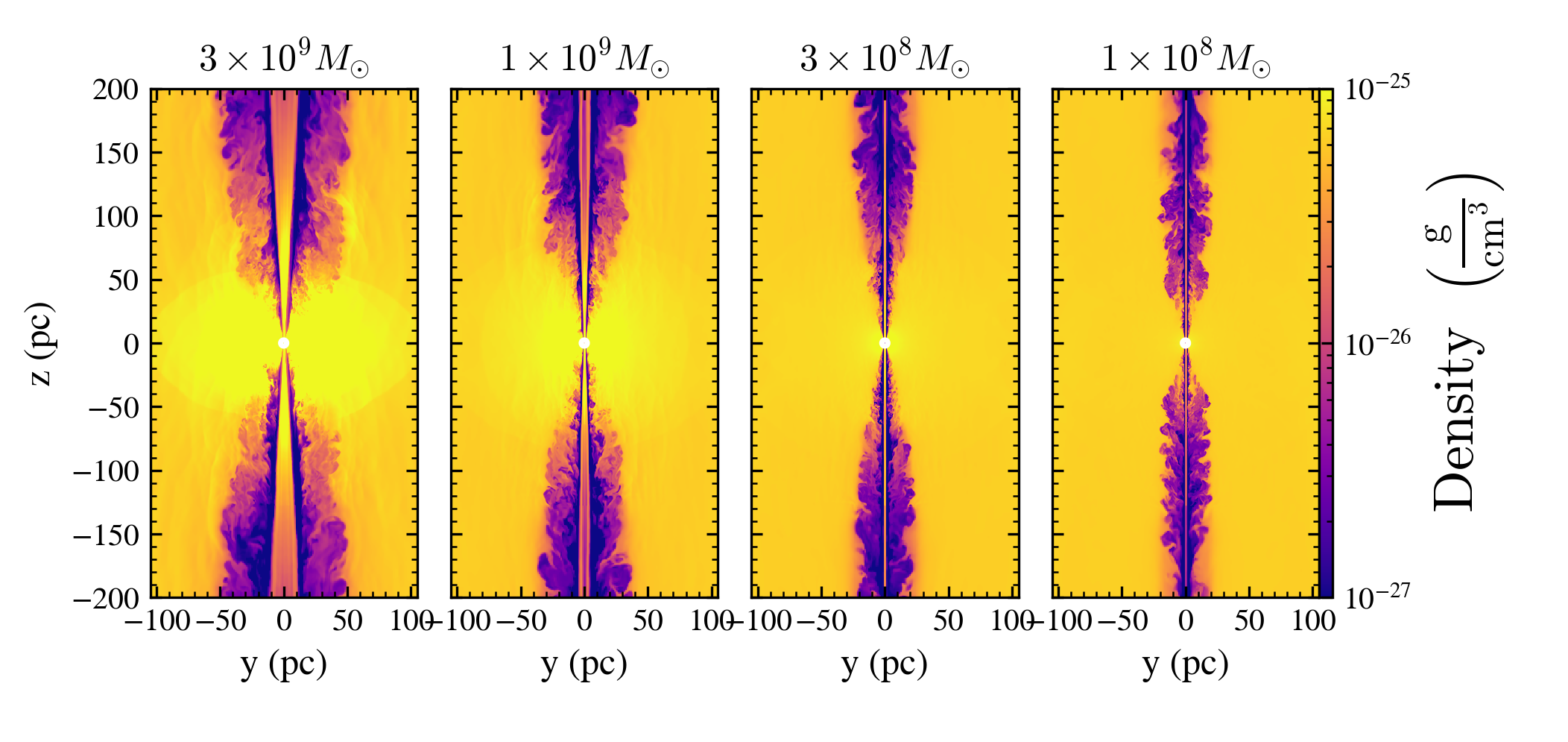}
    \caption{Gas density in the $yz$ plane at $t=0.2$~Myr for the JD runs, with fixed control surface size ($\rcs=2$~pc) and resolution ($\Delta x = 0.25$~pc) and fixed radius and height of the jet injection cylinder ($\rjet = \hjet = 0.5\ \rm pc$). The mass of the central black hole varies. }
    \label{slice_druns}
\end{figure*}

We tested the dependence of accretion and jet properties on the central SMBH mass via the set of JD runs. In these runs, we fix the parameters discussed above ($\rcs$, $\rdep$, $\rjet$ and $\hjet$) but vary the mass of the accretor, causing the relative proportion of these quantities to the Bondi radius to be different. In run JD4, the Bondi radius is around $7\Delta x$, which is smaller than the control surface radius. This provides an insight into how jet power and morphology change if the Bondi radius is less resolved when calculating the accretion rate.

Slice plots for the JD runs are shown in Figure~\ref{slice_druns}. With decreasing central SMBH mass (thus decreasing Bondi radius), the jet becomes less powerful and the shock region smaller. The lower mass SMBHs also sustain a lower energy budget, despite reaching a steady state. We see in Figures~\ref{ment} and \ref{profile2} that a lower-mass central SMBH generates less entrained material, and the gas also has lower velocity. This indicates the jet power is relatively smaller for lower mass SMBHs. We notice, however, comparing to the inflow gas in each individual run, the jet power is not necessarily small. In Figure~\ref{macc2} we observe that the accretion rate measured by the incoming flow is in fact higher for the lower mass SMBH compared to the Bondi-Hoyle-Lyttleton rate, indicating that when measuring accretion outside of the Bondi zone, we tend to over-estimate the accretion rate and thus the jet power. 

\subsubsection{Comparing with Bondi-Formula Accretion}

\begin{figure}
    \begin{center}
    \includegraphics*[width=1.\columnwidth]{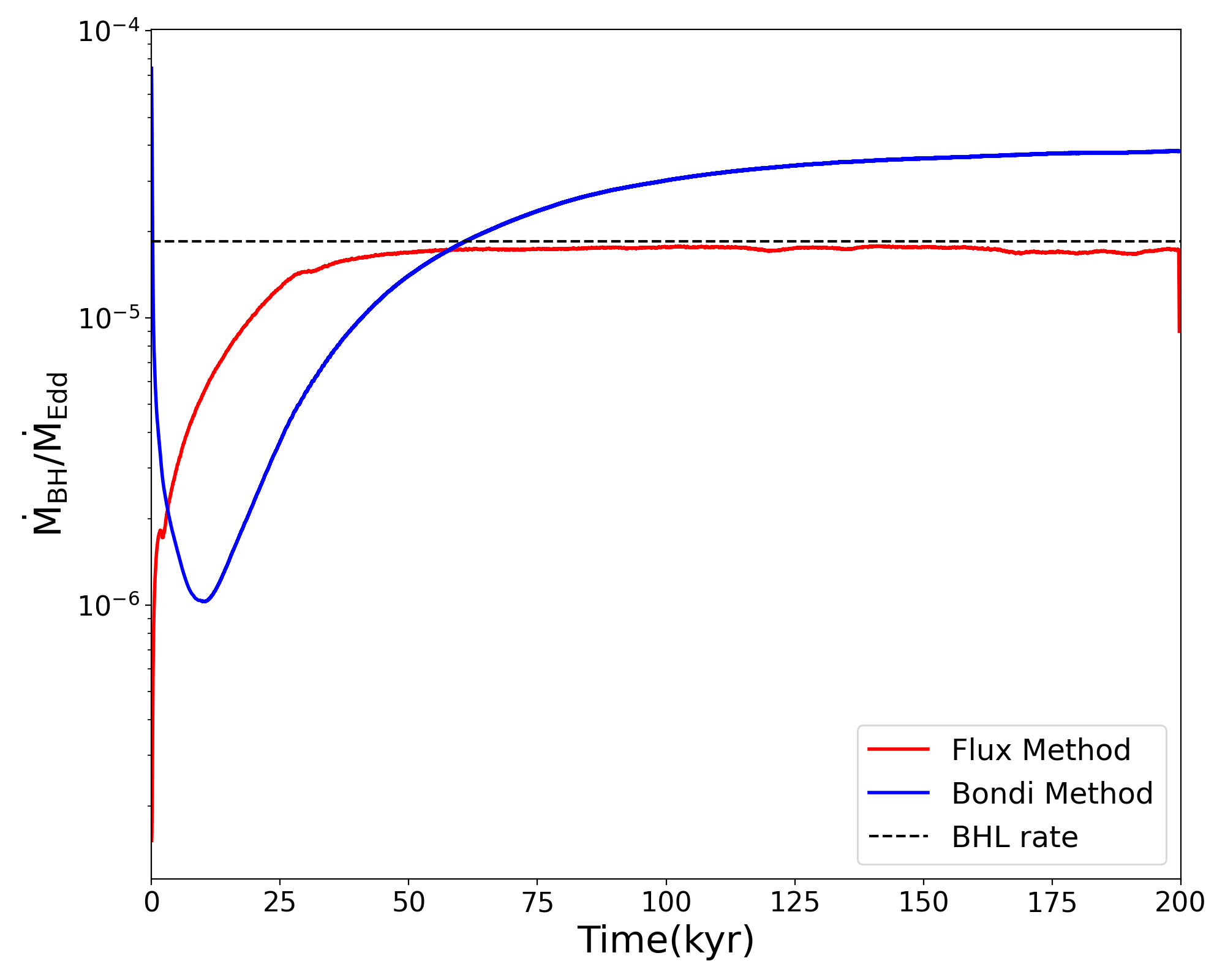}
    \caption{Evolution of accretion rate for different accretion scenarios (with measuring accretion rate using the Bondi formula or using our flux approach, run J0 vs.\ JC1 in Table~\ref{table1}), with the same initial condition and jet model. We use a fixed control surface size ($\rcs = 2$~pc), jet injection region size ($\hjet = 2~\Delta x, \rjet = 2~\Delta x$), and resolution ($\Delta x = 0.25$~pc).}
    \label{macc3}
    \end{center} 
\end{figure}

To compare our sink particle model with the traditional approach based on the Bondi formula, we ran a comparison run where the initial setup, jet model, boundary conditions, and reset criteria remain the same (using the most common configuration from run JC1), while only the method of accretion rate measurement is different. To determine the accretion rate in the ``Bondi formula'' run, we compute the sound speed and density by averaging zones lying within the control surface but not the jet injection cylinder. We still apply kernel resetting. To compare with similar simulations in the literature, we use the $\alpha$-Bondi configuration, where $\alpha = 100$, to calculate the final accretion rate and thus the jet power for injection. Figure~\ref{macc3} compares the effect of the different approaches on the evolution of the accretion rates. 

Using the control surface method produces a relatively steady accretion rate after an initial ramping up, while using the Bondi approach causes the accretion rate to initially drop before recovering to a value of about twice the expected rate. This suggests that the Bondi approach overestimates the initial accretion rate and thus the amount of feedback, blowing material away from the black hole and dropping the accretion rate temporarily. The control surface method only accounts for the fraction of solid angle not occupied by outflowing gas, so it allows the accretion rate to ramp up slowly even in the presence of a jet.

It is surprising that the control surface method reproduces the expected Bondi accretion rate while the Bondi method itself overestimates the accretion rate. The Bondi accretion solution does not account for outflow. Since outflow exists in the simulation, assuming that accretion rate must therefore overestimate the true accretion rate unless the outflow lowers the density or increases the sound speed within the averaging region. The control surface method permits the inflow in directions away from the jet to adjust itself in a way that somehow reproduces the Bondi rate. We would argue that the latter behavior is more likely to be physically correct, as it relies more on the resolved non-jet flows outside the control surface than on the barely resolved resetting region.

\begin{figure}
    \begin{center}
    \includegraphics*[width=1.\columnwidth]{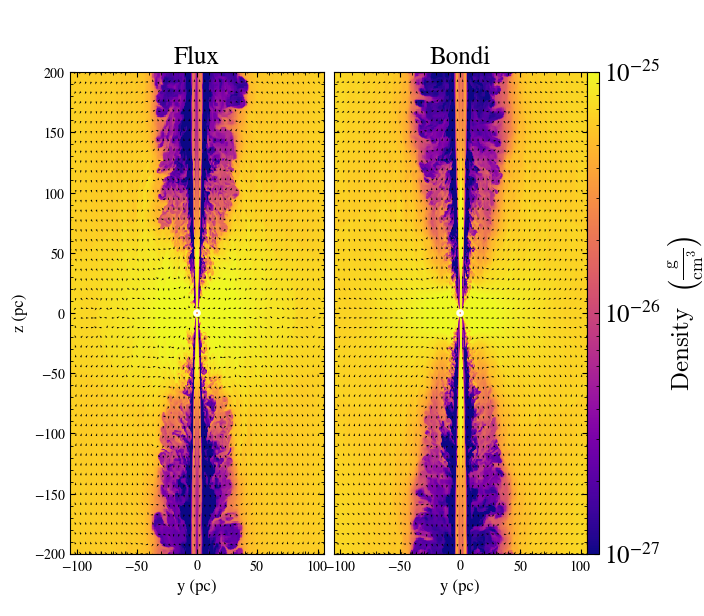}
    \caption{Density slices at same time ($t=0.2$~Myr) for runs with different accretion mechanisms (the runs shown are the same as Figure~\ref{macc3}). }
    \label{slice_jet_bondi}
    \end{center}
\end{figure}

The accretion behavior affects the structure of the jet, as seen in Figure~\ref{slice_jet_bondi}. Since the average accretion rate is larger for the Bondi case, the feedback energy deposited is also greater, so at a given late time in the run the head of the jet becomes wider and the central channel becomes more prominent. 

Our results are consistent with those of \cite{Negri}, who concluded that with the Bondi formalism in high resolution simulations, the BH accreted mass is generally overestimated. On the other hand, underresolved simulations underestimate the accreted mass; this is the reason for using a boost parameter. Due to the complex interplay between accretion and feedback, there is no reason to expect that the Bondi formula accurately predicts the accretion rate in realistic situations.

\section{Discussion and Comparison with Previous Work}

\subsection{Comparison with Smaller-Scale Jet Simulations}

To measure the convergence of our jet model, we compare its stability, collimation, and mass entrainment properties with jets we observe in parsec- and smaller-scale simulations.

The comparison with GRMHD simulations can be difficult, since these simulations usually do not account for any cooling, radiative processes, or interaction between jets and the surrounding medium. However, we can aim to build a qualitative picture of how accretion powers the launching of the jets. In GRMHD simulations, usually the accretion disks start at a low accretion rate as the magnetorotational instability builds up gradually inside the torus, and after a certain time scale (around $10^4\ r_g/c$\newtext{, where $r_g \equiv GM/c^2$}) the accretion rate steadily grows until it saturates, after which the accretion rate remains steady. This behavior is captured by our accretion model, where under the control surface criterion the accreted material takes time to build up enough energy to launch the jets. However, the steady state value for accretion rate in GRMHD simulations \citep[e.g.][]{tche_review} (usually rest-mass accretion rate) tends to be 10-100 times the Bondi accretion rate, due to the scale difference. 

As the jet propagates, we are also able to observe some signatures similar to MHD jets. In general, 3D relativistic, magnetized jet flows can exhibit many different kinds of instabilities \citep[for analysis of jet instabilities, see][]{Chow2023}. Instabilities such as current-driven instability and rotation-induced instability mainly occur in magnetically dominated jets when the current flow is high, which are beyond the scope of this work. However, the Kelvin-Helmholtz instability (KHI), which occurs at the contact discontinuity or the relatively-flowed shear layers, does not depend strongly on magnetic properties and is observed in our simulations. The KHI contributes to the formation of the small-scale turbulence effects at the jet-ICM interface. Through reconfinement, the jets also develop centrifugal instability (CFI) that generates turbulent flow \citep{Gourgoliatos}.

Many kinds of instabilities are seen in our jet runs, especially with short, \newtext{wide} jets, which is similar to what \cite{Monceau} has observed: a jet with a wider opening angle decelerates faster and produces a wider radial expansion zone dominated by instabilities. The Kelvin-Helmholtz instability is the most prominent since it is expected to occur in pure hydrodynamic simulations where there is a velocity shear.

Some other signatures of our simulated jets, including shock structure and energy transfer behavior, are similar to predictions from analytical models and previous simulations of the interaction of jets with the cluster-environment gas \citep[e.g.,][]{Begelman}. Our jets show a nose-cone structure as well as several other structures that represent various instabilities seen in MHD jet simulations \citep[e.g.,][]{Mignone}. Compared with works such as \cite{Perucho}, the ratio of the width of the jet shocked region to the distance from the BH ranges from about 200 (closer to the BH) to 20 (edge of the jet) in our simulations, which is similar to what they observed. The higher the jet power, the lower this ratio becomes, which is also consistent with the work mentioned above.

\subsection{Comparison and Implications for Larger Scale Cluster Simulations}

Based on our runs with different accretion and jet feedback configurations, we found certain differences with many galaxy cluster-scale simulations.

As indicated previously, most such simulations use the Bondi formula with a boost parameter to characterize accretion rates. Some models that aim to incorporate ``cold-mode accretion'' usually involve dividing the total amount of cold gas within the accretion zone by a certain timescale, or measuring the mass change within a few zones around the SMBH \citep[such as][]{Gaspari2013, Li}. This is also over-simplified, and involves arbitrary parameters that do not have a clear physical meaning and need to be tuned to observations. In our sink particle method, for the accretion model such parameters are taken out, at the cost of requiring higher spatial resolution in the vicinity of the AGN.

In comparison with bubble-type feedback models, our use of a cylindrical window function to impose feedback makes it easier to vary the shape and size of the jet; thus it can be accommodated to different sources and scenarios where the jet dynamics vary, and it is easier to incorporate effects such as precession. 

The main deficiency of the jet model presented here is that it still involves an ``efficiency parameter,'' which is set to be constant and consistent with previous AGN feedback work. We also do not distinguish hot and cold accretion fluxes on the control surface or take the angular momentum of the flow into account in the model.

\section{Conclusions}

In this work, we have developed a new subgrid model for accretion and jet feedback in hydrodynamic simulations. The model resembles a sink particle algorithm commonly used in star formation studies, measuring accretion onto SMBHs via a flux across a surface instead of using a simple analytical formula, and incorporating inner boundary conditions which remove the gas accreted. With this formulation we further investigated different ways to inject jets and characterize mechanical feedback and compared the jet dynamics and feedback properties with other smaller scale work to test the convergence of our model. We have found the following:

\begin{enumerate}
\item We tested the accretion part of the model with two idealized scenarios: spherically symmetric Bondi accretion and Bondi-Hoyle-Lyttleton accretion through a wind blowing past the BH particle. With a Bondi solution gas profile setup, we ran tests with varying resolutions, as well as varying the accretion measurement radius (control surface radius) and reset radius. We have managed to reproduce a steady-state flow as predicted with our sink particle model using a kernel reset that does not assume the Bondi solution, while maintaining an accretion rate within 0.1\% error compared to the Bondi accretion rate. 

\item The Bondi-Hoyle-Lyttleton test was conducted with varying Mach numbers for the ambient wind and different resolutions of $\rbhl$. As long as $\rbhl$ is resolved by two or more zones, we observe better than 10\% accuracy in the measured accretion rate. This means that for lower black hole masses or higher Mach numbers, the resolution requirements are more strict. To reliably reproduce the expected accretion rate within a reasonable error threshold, the control surface also needs to be at least two zones across. 

\item We further imposed a jet feedback model and studied the evolution of accretion and feedback as well as environmental effects the jet has on an idealized ICM environment. We ran simulations with different resolutions and jet sizes. Overall, properties including accretion rates, feedback energy, shocked volume, and amount of entrained mass show good convergence behavior with resolution. However, to reproduce some of the shock structures and fluid instabilities, a short, skinny jet injection region is favored. We adopt the $\rjet = \hjet = 1/4 \rcs$ configuration for future work.  

\item Our work has shown consistency with previous jet simulations at smaller scales, while as a standalone model it is able to reproduce the jet dynamics better than many current AGN feedback simulations have done. We are able to see some dynamic jet features that are observed in MHD jet simulations at parsec and kiloparsec scales, as well as instability at the jet-ICM interface. 
\end{enumerate}

This model still has several limitations. In the vicinity of the control surface, the numerical jet launching mechanism may interact with the incoming flow in ways that are not fully realistic. In addition, the feedback efficiency parameter described in jet models needs to take into account the accretion disk modeling we discuss in Paper~II. Relativistic effects are also not incorporated in the current jet model, but will likely impact the pressure of the shock inflated by the jet. 

The model will be used to address some of the key questions regarding cool-core clusters and the cool-core problem in later work. 

\begin{acknowledgments}
We acknowledge support from the US National Science Foundation under AAG 20-09868. FLASH was developed and is maintained largely by the DOE-supported Flash Center for Computational Science at the University of Chicago (now at the University of Rochester). Simulations were performed using Blue Waters at Illinois (ILL\_bawf, ILL\_bbbb) and Stampede2/3 at the Texas Advanced Computing Center (NSF ACCESS PHY230026).
\end{acknowledgments}

\software{FLASH \citep{Flash,Flash3},
          yt \citep{Turk.etal2011},
          Matplotlib \citep{Hunter.2007},
          NumPy \citep{harris2020array}
          }

\bibliographystyle{aasjournal}
\bibliography{references}

\end{document}